\documentclass[prb,reprint,superscriptaddress,twocolumn,amsmath,longbibliography]{revtex4-1}
\usepackage{hyperref}
\usepackage{graphicx}
\usepackage{natbib}

\usepackage{color}
\usepackage[normalem]{ulem} 

\newcommand\omegad{\omega_\mathrm d}
\newcommand\ee{\mathrm e}
\newcommand\integ[3]{\int_{#1}^{#2}\mathrm d#3\,}
\newcommand\ii{\mathrm i}
\renewcommand\phi{\varphi}
\renewcommand\rho{\varrho}
\newcommand\omegac{\omega_\mathrm c}
\DeclareMathOperator\sgn{sgn}
\DeclareMathOperator\im{Im}

\definecolor{DarkGreen}{rgb}{0,0.4,0}

\begin{document}

\title{Signature of the transition to a bound state in thermoelectric quantum transport}
\date{November 28, 2018 --- Modified: July 24, 2019}

\author{\'Etienne Jussiau}
\email{etienne.jussiau@lpmmc.cnrs.fr}
\affiliation{Universit\'e Grenoble Alpes, CNRS, LPMMC, 38000 Grenoble, France}
\author{Masahiro Hasegawa}
\email{h.masahiro@issp.u-tokyo.ac.jp}
\affiliation{Institute for Solid State Physics, The University of Tokyo, Kashiwa, Chiba 277-8581, Japan}
\author{Robert S. Whitney}
\email{robert.whitney@grenoble.cnrs.fr}
\affiliation{Universit\'e Grenoble Alpes, CNRS, LPMMC, 38000 Grenoble, France}

\begin{abstract}
We study a quantum dot coupled to two semiconducting reservoirs, when the dot level and the electrochemical potential are both close to a band edge in the reservoirs. This is modelled with an exactly solvable Hamiltonian without interactions (the Fano-Anderson model). The model is known to show an abrupt transition as the dot-reservoir coupling is increased into the strong-coupling regime for a broad class of band structures. This transition involves an infinite-lifetime bound state appearing in the band gap. We find a signature of this transition in the continuum states of the model, visible as a discontinuous behaviour of the dot's transmission function. This can result in the steady-state DC electric and thermoelectric responses having a very strong dependence on coupling close to critical coupling. We give examples where the conductances and the thermoelectric power factor exhibit huge peaks at critical coupling, while the thermoelectric figure of merit $ZT$ grows as the coupling approaches critical coupling, with a small dip at critical coupling. The critical coupling is thus a sweet spot for such thermoelectric devices, as the power output is maximal at this point without a significant change of efficiency.
\end{abstract}
\maketitle

\section{Introduction}
\begin{figure}[b]
\includegraphics[width=\linewidth]{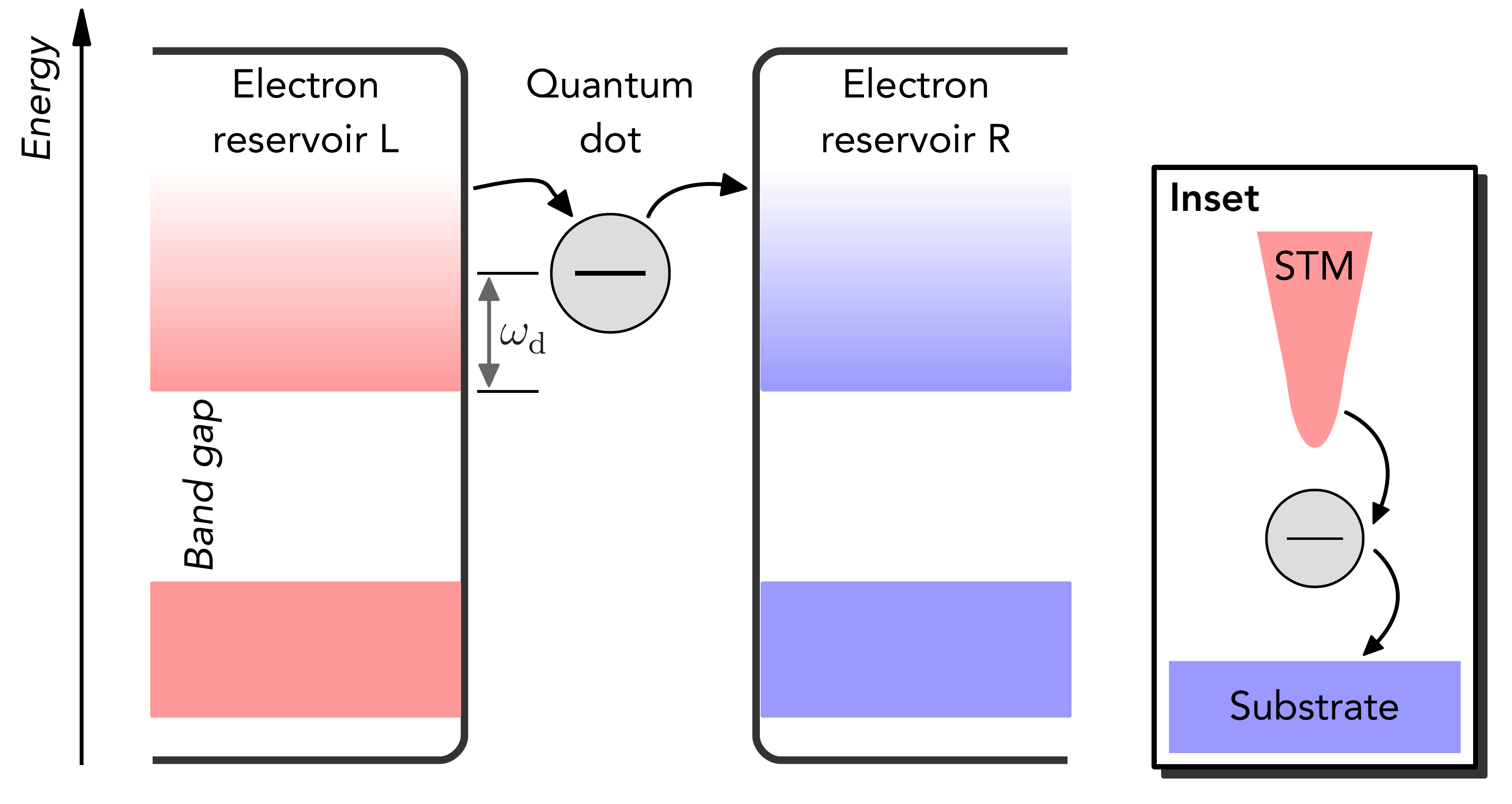}
\caption{A sketch of the type of system considered here, a single-level quantum dot between two semiconducting reservoirs. Inset: To study the system dynamics as a function of the dot-reservoir coupling, one can place the dot in an STM geometry (or a break junction) as shown.}
\label{Fig-System}
\end{figure}

There is great current interest in the thermal and thermoelectric transport properties of quantum dots or molecules coupled to electronic reservoirs. They can be used as heat engines (converting a heat flow into electrical power) or refrigerators (using electrical power to extract heat from an already cold reservoir of electrons). Experimental demonstrations include Refs.~[\onlinecite{Reddy2007Mar,Prance2009,Balachandran2012Aug,Molenkamp2015,Roche2015Apr,Hartmann2015Apr}], while much of the theory is reviewed in Ref.~[\onlinecite{ReviewBCSW}]. However, few such works have considered the effects of the band structure of the electronic reservoirs on the quantum dot's transport properties. We ask how the physics changes when the dot level is at an energy close to a band edge in semiconducting electronic reservoirs, see Fig.~\ref{Fig-System}. 
Here, as a first step in answering this question, we consider a non-interacting model, often known as the Fano-Anderson model,\cite{Anderson1961,Fano1961,BookMahan,BookCohen-Tannoudji,Topp2015} for which one can exactly solve the dynamics and extract all observables. We are particularly interested in the regime of strong coupling between the dot and reservoirs, see e.g.~Refs~[\onlinecite{Apertet2012Mar,Topp2015,Katz2016May,Whitney2018Aug,Strasberg2017Jun,Strasberg2018May,Dou2018Oct}], 
because one expects that stronger coupling allows for larger currents, and hence larger power output of quantum dot heat-engines.

This type of model has long been known to exhibit an infinite-lifetime state when the reservoir has a band edge. When the band edge is due to the reservoir's single particle band structure, the state is usually known as a \textit{bound state} or \textit{localised mode}. However, when it is due to a superconducting gap in the reservoir (which is not a situation we consider here) it is also known as a non-magnetic Shiba states.\cite{Shiba1973Jul} This bound state (or localised mode) was discussed in two textbooks,\cite{BookMahan,BookCohen-Tannoudji} and was greatly studied in bosonic models of an atomic level coupled to a photonic continuum with a band structure, initially for a zero-temperature continuum \cite{John1990, John1991,John1994,Kofman1994} reviewed in Refs.~[\onlinecite{Angelakis2004,Longhi2007May,Chang2018}], and more recently for finite-temperature.\cite{Zhang2012,Lei2012May,Xiong2015Aug,Ali2015Dec,Ali2017Mar}
It stops an atom's excited state fully decaying into the continuum, as recently observed in an NV centre in a waveguide,\cite{Liu2017} and is predicted to lead to perfect subradiance.\cite{Gonzalez-Tudela2017} The bound state (or localized mode) was extensively studied in the context of a quantum dot coupled to finite temperature fermionic reservoirs,\cite{Maciejko2006Aug,Dhar2006Feb,Stefanucci2007May,Jin2010Aug,Xiong2015Aug,Yang2015Oct,Tu2016Mar,Zhang2012,Engelhardt2016,Lin2016} exhibiting the same absence of decay, and even infinite-time oscillations. It also induces Landau-Zener-Stueckelberg physics in a slowly pumped dot.\cite{Basko2017}

The bound state's occupation depends on particle statistics and temperature, however its general properties (e.g.~whether it exists or not, its energy and wavefunction) depend only on the single-particle spectrum of the model Hamiltonian. If the continuum's density of states vanishes at the band edge, and the dot level is at an energy inside the band, then there is no bound state for weak dot-reservoir coupling. However, a bound state appears when the coupling exceeds a critical value.\cite{BookCohen-Tannoudji,Dhar2006Feb,Stefanucci2007May,Xiong2015Aug,Yang2015Oct,Zhang2012,Engelhardt2016,Lin2016} 

In this work, we investigate the changes in the \textit{continuum states} which accompany the appearance of a bound state (we refer readers to the works cited above for the physics of the bound states themselves). These changes in the continuum states can be observed via the transmission function of the system, which is crucial in determining the electrical and thermoelectric transport properties of the system. We show how the emergence of an infinite-lifetime bound state affects the transmission function and DC transport properties of a quantum dot coupled to two reservoirs.

One might guess that there would be little change in the continuum states at the transition, and hence little change in the 
 transmission and the currents. However, the transition involves an eigenstate emerging from the continuum, and it is natural that this cannot occur without some re-organisation of the continuum states. We show that the transition is accompanied by a singular change in the dot's transmission function at the band-edge. Hence, the transition leads to rapid changes in the electrical conductivity $G$, thermal conductivity $C$, and Seebeck coefficient $S$, when the electrochemical potential is close to the band edge.

This has consequences on the system's thermoelectric power factor $GS^2$ and dimensionless figure of merit $ZT$. 
We show that there is a huge peak in $GS^2$, which means a huge peak in the thermoelectric power output close to critical coupling. This is accompanied by a sharp small-amplitude variation in $ZT$. It means that maximal power generation in such cases is close to critical coupling, whereas the efficiency of the device does not vary significantly at this point. 
This thermoelectric response is very different from cases without band gaps where bound states cannot exist.\cite{BookSchaller,Topp2015}

Neither the bound state nor the transition are captured by the standard ``weak-coupling'' theory reviewed in Chapters 8 and 9 of Ref.~[\onlinecite{ReviewBCSW}]; this shows the importance of developing strong-coupling transport theories which capture the bound state physics. By doing this for a non-interacting system in this work, we hope to open the way to developing a transport theory which captures strong coupling effects like bound states in interacting systems.

\subsection{Contents of this work}

Section~\ref{Sect:model} introduces the model, section~\ref{Sect:laplace} reviews its exact solution, and section~\ref{Sect:general-currents} gives the general expression for current at time $t$. The central results of this work are in sections \ref{Sect:Long-time-currents}-\ref{Sect:thermoelectric}. Section \ref{Sect:Long-time-currents} gives the long-time current, showing that the DC currents take a Landauer form (once oscillations are neglected). Section~\ref{Sect:transmission} shows the transmission function has a discontinuous dependence on the dot-reservoir coupling when this coupling goes through its critical value. Section~\ref{Sect:lamb} gives a handwaving interpretation of these results in terms of a Lamb shift. Section~\ref{Sect:thermoelectric} shows the electric and thermoelectric response, with examples where some transport properties have huge peaks (while others have small dips) at the critical value of the coupling.

\section{The model}
\label{Sect:model}

The Hamiltonian that we study describes a single-level quantum dot coupled to two reservoirs,\cite{hbar-note}
\begin{equation}
\hat H=\omegad\hat d^\dagger\hat d+\sum_{\alpha,k}\omega_{\alpha k}\hat c_{\alpha k}^\dagger\hat c_{\alpha k}+\sum_{\alpha,k}\left(g_{\alpha k}\hat d^\dagger\hat c_{\alpha k}+g_{\alpha k}^*\hat c_{\alpha k}^\dagger\hat d\right)\!,
\label{H}
\end{equation}
where $\hat d$ and $\hat c_{\alpha k}$ denote field operators for an electron on the dot and in mode $k$ of reservoir $\alpha\in\{\mathrm L,\mathrm R\}$ respectively; the corresponding energies are $\omegad$ and $\omega_{\alpha k}$. Finally, $g_{\alpha k}$ describes the coupling between the dot and mode $k$ in reservoir $\alpha$. This model neglects electron-electron interactions on the dot. The simplest experimental implementation of such a model is to consider an interacting quantum dot (described by an Anderson impurity Hamiltonian) with a large enough magnetic field that the dot's spin-state with higher energy is always empty, which makes the on-dot interaction term negligible. The electron reservoirs contain infinitely many modes described by continuous spectral densities
\begin{equation}
J_\alpha(\omega)=\sum_{\alpha,k}|g_{\alpha k}|^2\delta(\omega-\omega_{\alpha k}).
\end{equation}
Crucially, we do \textit{not} take the wide-band limit, and instead consider the case where the dot level is close to a band edge. If not otherwise specified, the results in this work are for an arbitrary band structure in the reservoirs (possibly with multiple bands and band gaps). However, we will particularly consider reservoirs featuring a single band, with the dot level close to the lower band edge. We then consider that the reservoir's spectral density goes like a power law with exponent $s$ at this band edge, and is regularised with an exponential cut-off at high energies. We restrict our interest to two reservoirs made of the same material in the linear-response regime,\cite{electroneutrality-note} so
\begin{equation}
J_\mathrm L(\omega)=J_\mathrm R(\omega)=
\begin{cases}
\displaystyle\frac K2{\left(\frac\omega\omegac\right)\!}^s\ \ee^{-\omega/\omegac}
&\text{for $\omega>0,$}\\
0&\text{for $\omega<0$,}
\end{cases}
\label{J}
\end{equation}
where (without loss of generality) we take the zero of energy to be the band edge. 

\subsection{Bound states}

This Hamiltonian exhibits bound states whose energy and wavefunction are independent of statistics of the particles (fermionic or bosonic) and reservoir temperatures. The simplest way to see this is to considers the single-particle spectrum of Eq.~\eqref{H} for a finite number of reservoir modes. Because it is quadratic, a Bogoliubov transformation can be performed. Then $\hat H$ takes the form of an arrowhead matrix whose eigenvalues are given in Ref.~[\onlinecite{OLeary1990}]. Upon taking the number of reservoir modes to infinity, we find that the discrete eigenmodes (not in the continuum) must have energy $\omega_{*n}$ satisfying the equation\cite{OLeary1990,BookMahan,BookCohen-Tannoudji} 
$\Omega(\omega_{*n})=0$,
 where 
\begin{equation}
\Omega(\omega)=\omega-\omegad-\integ{}{}{\omega'}\frac{J(\omega')}{\omega-\omega'},
\label{Omega}
\end{equation}
at the same time as $\omega_{*n}$ being in a band gap; $J(\omega_{*n})=0$. In band gaps, $\Omega(\omega)$ grows monotonically with $\omega$, so there is at most \textit{one} bound state in each band gap. Expressions for the eigenvector corresponding to such an eigenvalue can also be found. These eigenmodes of Eq.~\eqref{H} are called bound states (or localized modes) because they do not decay with time.

\begin{figure}
\includegraphics[width=0.6\linewidth]{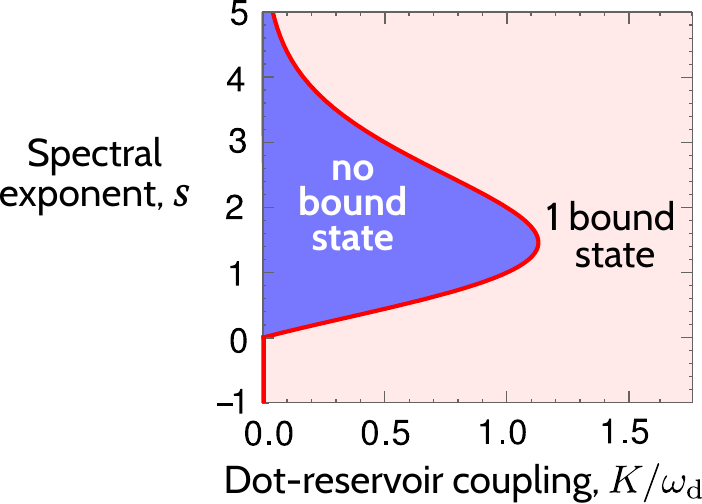}
\caption{Diagram\cite{Zhang2012,Lin2016} of the two possible regimes for a system with the spectrum in Eq.~\eqref{J}, as a function of the coupling strength $K$ and the spectral exponent $s$. For given $s$, a bound state appears when the coupling $K$ exceeds its critical value $K_*$ (red curve) given by Eq.~\eqref{K_*-our-spectrum}.}
\label{Fig-Transition}
\end{figure}

For the reservoir spectrum with a single band gap (for all $\omega<0$) in Eq.~\eqref{J}, $\Omega(\omega)$ monotonically grows with $\omega$ in the band gap. As $\Omega(-\infty)=-\infty$, there is one bound state in the band gap if $\Omega(\omega\to0^-)>0$, and none otherwise. This condition is satisfied if $K>K_*$, where the critical coupling\cite{Zhang2012,Lin2016}
\begin{equation}
K_*=
\begin{cases}
\ \ 0&\text{for $s\le0$,}\\
\ \omegad/\Gamma(s)&\text{for $s>0$.} 
\end{cases}
\label{K_*-our-spectrum}
\end{equation}
This is sketched in Fig.~\ref{Fig-Transition}. When the spectral exponent $s$, is negative (e.g.~a divergence in $J(\omega)$ at the band edge) the bound state is always present. In contrast, for all $s >0$, the critical coupling $K_*$ is finite and of the order of the energy difference $\omegad$ between the dot level and the band edge, so the transition will be observed in systems with the dot level close to the band edge. This is very different from the wide-band limit, which assumes that the dot level is infinitely far from any band edge, so there cannot be any bound state at finite coupling.

Returning to the general case, we note that while a bound state's existence, energy and wavefunction do not depend on the particle statistics or temperature, the occupation of bound states (and the occupation of the continuum states) depend on both statistics and temperature. Thus to study observables such as the dot occupation or the current through the dot, this exact diagonalisation is not sufficient. It is also inconvenient, and we prefer to use Heisenberg equations of motion to find the time dependence of such 
observables.

These bound states should not be confused with the ``dark-state'' that forms in a multi-level system coupled to a wide-band continuum, when two of the system's levels are degenerate.\cite{Ping2011Apr,Ping2013Mar,Xu2014Aug}

\section{Solution via Laplace transform of equations of motion}
\label{Sect:laplace}

The Hamiltonian in Eq.~\eqref{H} is quadratic, so it can be solved using many different methods; such as Heisenberg equations of motion, \cite{Fano1961,John1994,Kofman1994,Ping2011Apr,Zhang2012,Angelakis2004,Xu2014Aug,Topp2015,Xiong2015Aug,Engelhardt2016,Ali2017Mar,Chang2018} Feynman-Vernon path integrals, \cite{Jin2010Aug} extended quantum Langevin equations,\cite{Yang2015Oct} Green's functions,\cite{Anderson1961} and Keldysh Green's functions.\cite{Schiller1998Dec,Maciejko2006Aug,Stefanucci2007May,Zhang2012,Tu2016Mar} Here we use the Heisenberg equations of motion, which consist of a set of linear first-order differential equation,\cite{Fano1961,John1994,Kofman1994,Ping2011Apr,Zhang2012,Angelakis2004,Xu2014Aug,Topp2015,Xiong2015Aug,Engelhardt2016,Ali2017Mar,Chang2018} solved using a Laplace transform. Let $\hat D(z)$ and $\hat C_{\alpha,k}(z)$ be the Laplace transforms of $\hat d(t)$ and $\hat c_{\alpha k}(t)$, where the operators are time-dependent because we work in the Heisenberg picture, e.g.~$\hat D(z)=\integ0\infty t\ee^{zt}\hat d(t)$. Then
\begin{align}
&\hat D(z)=\frac1{z+\ii(\omegad+\Sigma(z))}\left(\hat d_0-\ii\sum_{\alpha,k}\frac{g_{\alpha k}\hat c_{\alpha k;0}}{z+\ii\omega_{\alpha k}}\right),\label{Laplace_solution_D(z)}\\
&\hat C_{\alpha k}(z)=\frac{1}{z+\ii\omega_{\alpha k}}\left(\hat c_{\alpha k;0}-\ii g^*_{\alpha k} \hat D(z)\right),\label{Laplace_solution_C(z)}
\end{align}
where $\hat d_0$ and $\hat c_{\alpha k;0}$ are the operators at time $t=0$. The prefactor in Eq.~\eqref{Laplace_solution_D(z)} contains the self-energy
\begin{equation}
\Sigma(z)=\sum_{\alpha,k}\frac{|g_{\alpha k}|^2}{\ii z-\omega_{\alpha k}}=\integ{}{}\omega\frac{J(\omega)}{\ii z-\omega},
\label{Sigma}
\end{equation}
where $J(\omega)=J_\mathrm L(\omega)+J_\mathrm R(\omega)$. For $J(\omega)$ in Eq.~\eqref{J},
\begin{equation}
\Sigma(z)=-K\Gamma(1+s){\left(-\frac{\ii z}{\omegac}\right)\!}^s\ \Gamma\left(-s,-\frac{\ii z}{\omegac}\right)\ee^{-\ii z/\omegac},
\end{equation}
where $\Gamma(a)$ and $\Gamma(a,w)$ respectively denote the complete and incomplete Gamma functions.

This is a complete formal solution of the problem, but it requires an inverse Laplace transform to find its consequences for a given observable. This inverse Laplace transform\cite{BookAppel} is a contour integral of the type sketched in Fig.~\ref{Fig-Contour}b (see Appendix \ref{Sect:inverse-laplace}). Eq.~\eqref{Laplace_solution_D(z)} is a product of two terms, so its inverse Laplace transform is a convolution in the time domain of the inverse Laplace transform of each term. We define $\phi(t)$ as the inverse transform of the prefactor, $1/(z+\ii(\omegad+\Sigma(z)))$, this was calculated in Ref.~[\onlinecite{Zhang2012}] (see our Appendix~\ref{Sect:inverse-laplace}) which showed that it contains physics of both the continuum and the bound states. By comparison, the inverse transform of the other terms in Eqs.~(\ref{Laplace_solution_D(z)},\ref{Laplace_solution_C(z)}) are relatively straightforward.

\begin{figure}
\includegraphics[width=0.95\linewidth]{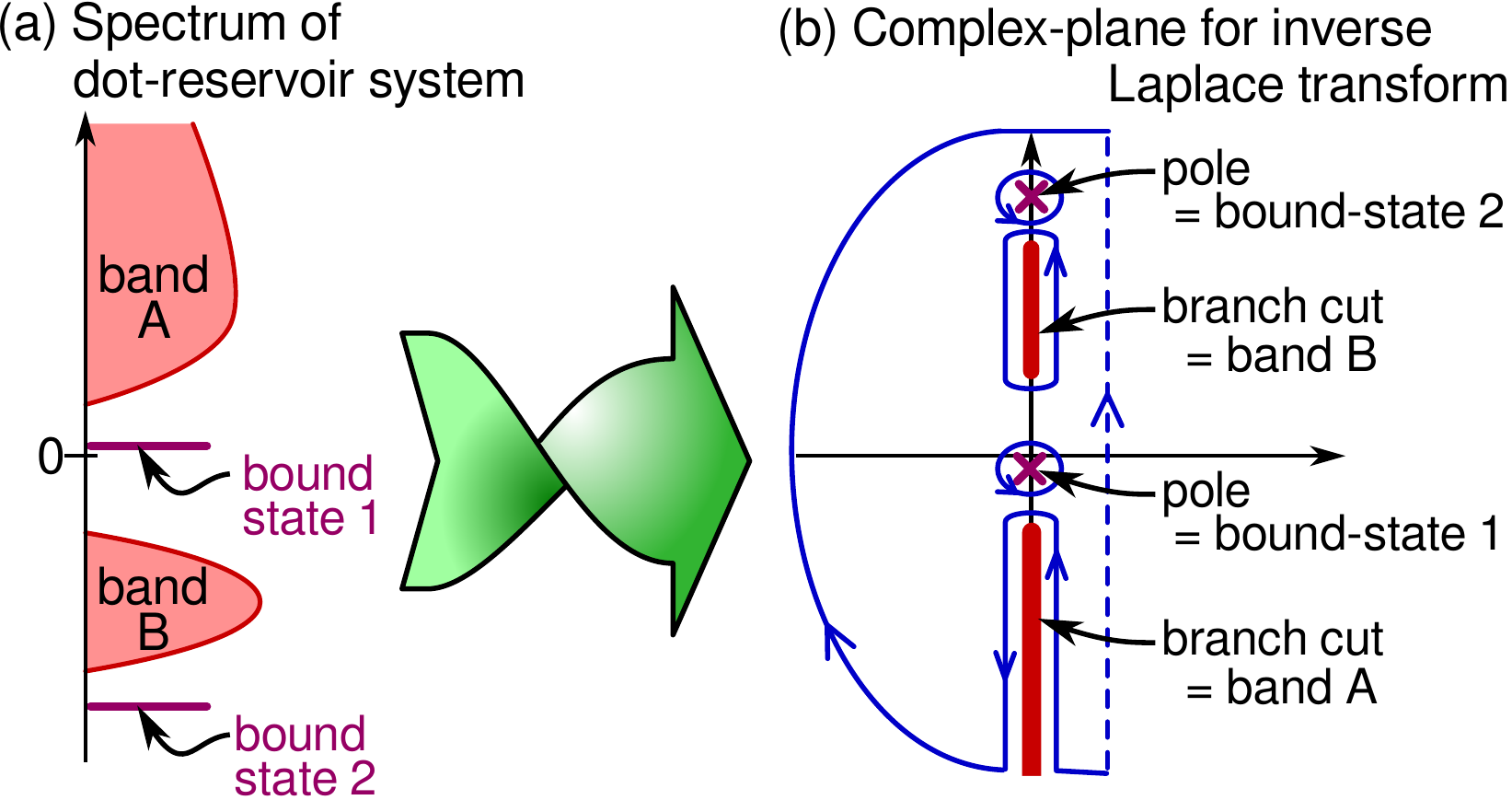}
\caption{An example\cite{Zhang2012} of the correspondence between (a) the spectrum of Eq.~\eqref{H}, and (b) structures on the imaginary axis in the inverse Laplace transform $\phi(t)$. A bound state at energy $\omega_{*n}$ corresponds to a pole in the complex plane at $z=-\ii\omega_{*n}$. A band from energy $\omega$ to energy $\omega'$ corresponds to a branch cut from $z=-\ii\omega$ to $z=-\ii\omega'$. The inverse Laplace transform is an integral along the dashed blue contour in (b), deformed into the solid blue contour for evaluation.\cite{BookAppel}
}
\label{Fig-Contour}
\end{figure}

\section{General results for currents}
\label{Sect:general-currents}

This section briefly summarizes what is known\cite{Lei2012May,Jin2010Aug,Yang2015Oct} about the time-dependent currents in this model. The particle current into reservoir $\alpha$ is defined as 
\begin{equation}
j_\alpha^{(\mathrm N)}(t)=\frac{\mathrm d}{\mathrm dt}\sum_k\,\langle\hat c_{\alpha k}^\dagger(t)\hat c_{\alpha k}(t)\rangle.
\end{equation}
The energy current entering reservoir $\alpha$ is defined as
\begin{equation}
j_\alpha^{(\mathrm E)}(t)=\frac{\mathrm d}{\mathrm dt}\sum_k\,\omega_{\alpha k} \ \langle\hat c_{ \alpha k}^\dagger(t)\hat c_{\alpha k}(t)\rangle.
\end{equation}
These current can be found by substituting in the inverse Laplace transforms of Eqs.~(\ref{Laplace_solution_D(z)},\ref{Laplace_solution_C(z)}) and taking the expectation value with respect to the initial state (the state at $t=0$). Here, we consider the initial state to be a product state between the dot and the reservoirs, with the dot in a chosen state and the reservoirs in a thermal state. This is natural when the system and reservoirs are initially decoupled, and we instantaneously turn on the coupling (a quench) at time $t=0$. Then the state at the moment of the quench ($t=0$) is
\begin{equation}
\hat\rho (t=0)=\hat p_0\otimes \hat\rho_\mathrm{L,eq} \otimes \hat\rho_\mathrm{R,eq},
\label{Eq:product-state}
\end{equation}
where $\hat p_0$ is the initial density matrix of the dot and $\hat\rho_{\alpha,\mathrm{eq}}$ is the equilibrium density matrix for reservoir $\alpha$,
\begin{equation}
\hat\rho_{\alpha,\mathrm{eq}}=\frac{\ee^{-\beta_\alpha \sum_k(\omega_{\alpha k}-\mu_\alpha)\hat c_{\alpha k;0}^\dagger\hat c_{\alpha k;0}}}{\mathrm{Tr}\left(\ee^{-\beta_\alpha \sum_k(\omega_{\alpha k}-\mu_\alpha)\hat c_{\alpha k;0}^\dagger\hat c_{\alpha k;0}}\right)},
\end{equation}
with $\beta_\alpha$ and $\mu_\alpha$ denoting the inverse temperature and the chemical potential of reservoir $\alpha$.

Then the particle current at time $t$ is\cite{Lei2012May,Jin2010Aug,Yang2015Oct} 
\begin{equation}
j_\alpha^{(\mathrm N)}(t)=2\im\bigg(
\begin{aligned}[t]
&n_0\phi^*(t)\integ{}{}\omega J_\alpha(\omega)\psi(t,\omega)\\
&+\integ{}{}\omega J_\alpha(\omega)f_\alpha(\omega)\psi^*(t,\omega)\ee^{-\ii\omega t}\\
&+\integ{}{}\omega\!\mathrm d\omega'\,
\begin{aligned}[t]
&J_\alpha(\omega)J(\omega')F(\omega')\\
&\times\psi^*(t,\omega')\chi(t,\omega,\omega')\bigg).
\end{aligned}
\end{aligned}
\label{j_alpha(t)}
\end{equation}
Here, $F(\omega)=\sum_\alpha J_\alpha(\omega)f_\alpha(\omega)/J(\omega)$, so it is the ``average'' of the Fermi functions $f_\mathrm L(\omega)$ and $f_\mathrm R(\omega)$. The time-dependent functions $\psi(t,\omega)$ and $\chi(t,\omega,\omega')$ are completely determined by $\phi(t)$,
\begin{align}
&\psi(t,\omega)&=&\ii\integ0tt'\phi(t')\ee^{-\ii\omega(t-t')}\\
&\chi(t,\omega,\omega')&=&\ii\integ0tt'\psi(t',\omega')\ee^{-\ii\omega(t-t')}.
\end{align}
The energy current $j_\alpha^{(\mathrm E)}(t)$ into reservoir $\alpha$ is the same as Eq.~\eqref{j_alpha(t)} with an extra factor of $\omega$ in each integrand.

Hence, the currents at time $t$ are determined by integrals of the function $\phi(t)$ calculated in Ref.~[\onlinecite{Zhang2012}], see our Eq.~\eqref{phi(t)}. The time-dependences of such currents were plotted in Refs.~[\onlinecite{Lei2012May,Jin2010Aug,Yang2015Oct}].

\section{Long-time limit of currents}
\label{Sect:Long-time-currents}

Here we analysis in detail the current in the long-time limit, making use of the continuity equation to get simple expressions for the current. The Riemann-Lebesgue lemma\cite{BookAppel} states that an integrable function's Fourier transform vanishes at infinity,\cite{Kofman1994,Xiong2015Aug} so for $\phi(t)$ given in Eq.~\eqref{phi(t)},
\begin{equation}
\phi(t\to\infty) = \sum_n Z_{*n} \,\ee^{-\ii\omega_{*n}t},
\label{phi_long-time}
\end{equation}
where Eq.~\eqref{Z_*-definition} gives $Z_{*n}$, defined as the overlap between the dot level and the bound state at energy $\omega_{*n}$.This allows us to find the long time limits of $\psi(t,\omega)$ and $\chi(t,\omega,\omega')$. The long-time current is the sum of a DC (time-independent) component and an oscillating component,
\begin{equation}
j_\alpha^{(\mathrm N)}(t\to \infty)=j_{\alpha;\mathrm{DC}}^{(\mathrm N)}(t\to\infty)
+j_{\alpha;\mathrm{osc}}^{(\mathrm N)}(t\to \infty).
\label{j-long-t}
\end{equation}
The DC component is
\begin{multline}
j_{\alpha;\mathrm{DC}}^{(\mathrm N)}(t\to\infty)\\
=\integ{\mathrm B}{}\omega J_\alpha(\omega)\big(J(\omega)A(\omega)F(\omega)-S(\omega)f_\alpha(\omega)\big),
\label{j_DC}
\end{multline}
where the subscript $\mathrm B$ indicates that the integral is over energies in the bands; i.e.~it is over those $\omega$ where $J(\omega)\neq 0$. Here we have introduced the dot's local density of states $S(\omega)$ given in Eq.~\eqref{S}. Finally, we have regrouped numerous terms in
\begin{equation}
A(\omega)=\left(\sigma(\omega)+ \sum_n\frac{Z_{*n}}{\omega-\omega_{*n}}\right)^2+\pi^2S^2(\omega)
\end{equation}
where $\sigma(\omega)$ is the following principal value integral
\begin{equation}
\sigma(\omega)=P\hskip-3.75mm\integ{}{}{\omega'}\frac{S(\omega')}{\omega-\omega'}.
\end{equation}

Defining $\omega_{nm}=\omega_{*n}-\omega_{*m}$, the oscillating component is given by the double-sum over bound states,
\begin{equation}
j_{\alpha;\mathrm{osc}}^{(\mathrm N)}(t\to\infty)=-\sum_{n,m} \,M_{nm} \,\Lambda_{\alpha;nm} \sin(\omega_{nm}t),
\label{j_osc}
\end{equation}
with $\Lambda_{\alpha;nm}=\Lambda_\alpha(\omega_{*n})-\Lambda_\alpha(\omega_{*m})$, where $\Lambda_\alpha(\omega)$ is the Lamb shift induced by reservoir $\alpha$ given by Eq.~\eqref{Lambda}, and 
\begin{equation}
M_{nm}=Z_{*n}Z_{*m}\left(n_0+ \integ{\mathrm B}{}\omega \frac{J(\omega)F(\omega)}{(\omega-\omega_n)(\omega-\omega_m)}\right).
\label{M_nm}
\end{equation}
The double-sum is zero unless there are two or more bound states. When it is non-zero, it oscillates at a frequency $\omega_{nm}$, as observed in Ref.~[\onlinecite{Yang2015Oct}]. To understand this, one should observe that if there are two or more bound states, then the dot occupation $n(t)$ oscillates forever\cite{Xiong2015Aug,Yang2015Oct,Ali2017Mar} at frequency $\omega_{nm}$, see Eq.~\eqref{n_long-time}, with a fraction of an electron moving from dot to reservoirs and back during each oscillation. This is the physical origin of the oscillating current $j_{\alpha;\mathrm{osc}}^{(\mathrm N)}(t\to\infty)$.

To simplify the result for the current in Eqs.~(\ref{j-long-t}-\ref{M_nm}), we use the continuity equation. This is similar in spirit to the simplification in Refs.~[\onlinecite{Caroli1971,Meir1992}], but is a little more involved because we allow for an oscillating dot state. The Heisenberg equations of motion for the Fano-Anderson model give rise to a continuity equation ensuring particle conservation,
\begin{equation}
\frac{\mathrm d}{\mathrm dt}n(t)=j_\mathrm L^{(\mathrm N)}(t)+j_\mathrm R^{(\mathrm N)}(t)
\label{cont-eq}
\end{equation}
where $n(t)$ is the dot occupation at time $t$ (see Appendix~\ref{Sect:dot-occupation}). The time-derivative of $n(t)$ solely consists of an oscillating contribution from the bound states. In the long-time limit, we find that it is exactly the same as the oscillating component of the currents in Eq.~\eqref{j_osc}, so the equality is automatically enforced for the oscillating terms. Therefore, to satisfy Eq.~\eqref{cont-eq}, we also need that the total DC current vanishes, using this with Eq.~\eqref{j_DC}, we find that
\begin{equation}
A(\omega)= S(\omega)\big/J(\omega).
\label{A-result}
\end{equation}
So we can replace the ugly function $J(\omega)A(\omega)$ by the local density of states $S(\omega)$. This is not only algebraically simpler (and so easier to calculate numerically), it also has a much clearer physical meaning than $J(\omega)A(\omega)$. 
The local density of states $S(\omega)$ inside the band is independent of the bound states (unlike $A(\omega)$)
which makes it explicitly clear that only the continuum states carry DC current. The long-time particle current in Eqs.~(\ref{j-long-t}-\ref{M_nm}) eventually becomes
\begin{multline}
j_\alpha^{(\mathrm N)}(t\to\infty)=\int\frac{\mathrm d\omega}{2\pi}\,\mathcal T(\omega)(f_{\overline\alpha}(\omega)-f_\alpha(\omega))\\
-\sum_{n,m} \, M_{nm} \, \Lambda_{\alpha;nm}\, \sin (\omega_{nm}t).
\label{j^(N)_steady}
\end{multline}
where $\overline\alpha=\mathrm L$ if $\alpha=\mathrm R$ and vice-versa. The integral corresponds to the continuum contribution, and it takes a Landauer form\cite{Landauer1957,Landauer1970,Meir1992,Topp2015} with the transmission function
\begin{equation}
\mathcal T(\omega)=\frac{4\pi^2 \, J_\mathrm L(\omega)\,J_\mathrm R(\omega)}{(\omega-\omegad-\Lambda(\omega))^2+\pi^2J(\omega)^2}\, ,
\label{transmission}
\end{equation}
for all $\omega$ where $J_\mathrm{L,R}(\omega)\neq 0$. The transmission function is zero for any $\omega$ where $J_\mathrm{L,R}(\omega)=0$. Here $\Lambda(\omega)$ is the total Lamb shift, $\Lambda(\omega)=\Lambda_\mathrm L(\omega)+\Lambda_\mathrm R(\omega)$, 
given by Eq.~\eqref{Lambda}.

The bound states' contribution to the long-time current corresponds to its oscillatory component given by the double sum in Eq.~\eqref{j^(N)_steady} Note that Eq.~\eqref{j^(N)_steady} differs from the well-known Meir-Wingreen formula,\cite{Meir1992} which does not contain this double sum; Appendix~\ref{Sect:adiabatic} explains that this is due to different initial conditions (initial quench versus adiabatic preparation). In electronic systems, these oscillations may be too small to measure and one typically only measures the steady-state DC currents. Furthermore, here we consider a model which never has two bound state, so there are no oscillations at long times. Thus we drop these oscillations and get the Landauer form\cite{Landauer1957,Landauer1970,Meir1992,Topp2015} for the steady-state DC particle current from L to R,
\begin{equation}
j_\mathrm{DC}^{(\mathrm N)}=\int\frac{\mathrm d\omega}{2\pi}\,\mathcal T(\omega)(f_\mathrm L(\omega)-f_\mathrm R(\omega)).\qquad
\label{j^(N)_dc}
\end{equation}
With algebraic manipulations similar to those for the particle current, we find that the steady-state DC energy current from $\mathrm L$ to $\mathrm R$ also takes a Landauer form,
\begin{equation}
j_\mathrm{DC}^{(\mathrm E)}=\int_{-\infty}^\infty\frac{\mathrm d\omega}{2\pi}\,\omega\, \mathcal T(\omega)(f_\mathrm L(\omega)-f_\mathrm R(\omega)).
\label{j^(E)_dc}
\end{equation}

It is crucial to note that the DC currents in Eqs~(\ref{j^(N)_dc},\ref{j^(E)_dc}) do not depend on the initial state of the dot, whether or not there is a bound state. This is very different from the dot occupation, which remembers its initial state forever if there is a bound state,\cite{Xiong2015Aug,Yang2015Oct,Ali2017Mar} see Fig.~\ref{Fig-nsteady} in Appendix~\ref{Sect:dot-occupation}. This difference comes from the fact that the bound state does not contribute to the steady-state DC currents. Hence, one might guess that the sudden emergence of the bound state at critical coupling ($K=K_*$) will not manifest itself in the DC current. However, we will show that, when the bound state emerges from the continuum, it is accompanied by an abrupt change in the continuum states (which carry the steady-state DC currents). We will show this by highlighting the discontinuity in the $K$ dependence of the dot's transmission function at $K=K_*$, which can have a strong effect on the dot's electric and thermoelectric transport properties.

\begin{figure}
\includegraphics[width=\linewidth]{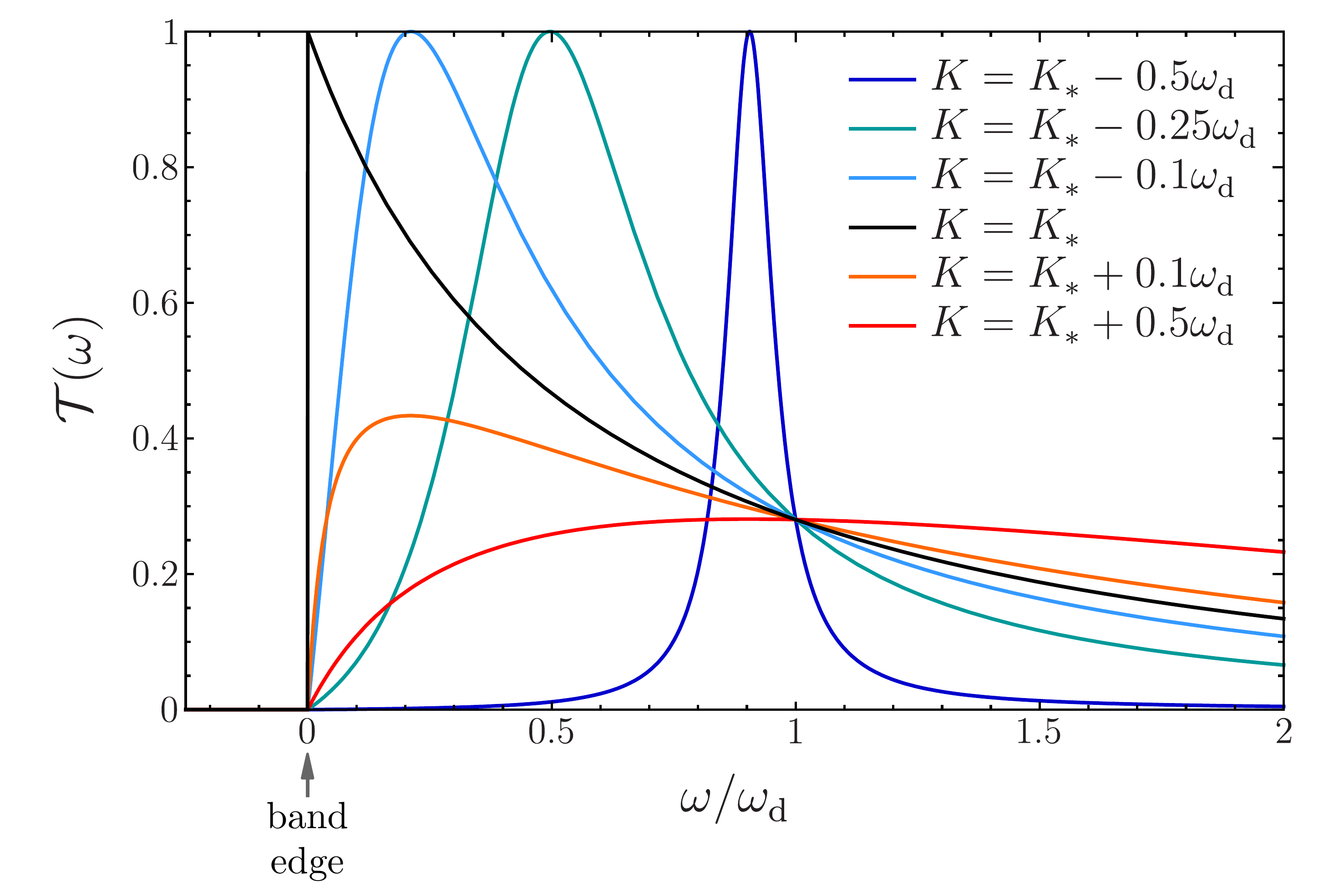}
\caption{Typical behaviour of the transmission as a function of energy for various values of the coupling, when $J(\omega)$ is given by Eq.~\eqref{J} and vanishes at the band edge in a sublinear manner ($0<s<1$). The parameters are $s=1/2$ and $\omegac=10\omegad$. For $K$ much less than $K_*$ (blue lines), the transmission function is a Lorentzian at the dot level. It then loses its shape and drifts towards the origin as the coupling increases up to its critical value (black line). When $K$ exceeds $K_*$ (orange lines), the transmission becomes much flatter.}
\label{Fig-T_s0p5}
\end{figure}

\section{Transmission function}
\label{Sect:transmission}

This section discusses the abrupt change in the transmission function, due to the abrupt changes in the continuum states which accompany the emergence of the bound state from the continuum. We consider the spectrum in Eq.~\eqref{J} with different values of the spectral exponent $s$. 

Intriguingly, despite its discontinuous behaviour at $K=K_*$ (which we will discuss in more detail below), one finds that the transmission at energy $\omega=\omegad$ is completely independent of the coupling $K$. This is most clearly visible as the point where all the curves cross in Fig.~\ref{Fig-T_s0p5}, but it is true for all $s$. The reason is that when $\omega=\omegad$, Eq.~\eqref{transmission} reduces to $\mathcal T(\omegad)=4\pi^2 \, J_\mathrm L(\omegad)\,J_\mathrm R(\omegad) \big/\big(\Lambda^2(\omegad)+\pi^2J(\omegad)^2\big)$. As both $J_\alpha(\omega)$ and $\Lambda(\omega)$ scale like $K$, one sees that $K$ cancels out of the expression for $\mathcal T(\omegad)$, so the transmission at this particular energy does not change with coupling.

\subsection{Transmission for $0<s<1$}
For the spectrum in Eq.~\eqref{J} with $0<s<1$ the transmission as a function of energy is qualitatively similar to that shown for $s=1/2$ in Fig.~\ref{Fig-T_s0p5}. The transition can be clearly seen as a discontinuity of the transmission at the band edge, $\mathcal T(\omega \to 0^+)$, as a function of $K$. The nature of the discontinuity at small $\omega$ depends on the exponent $s$. For $0<s<1$, the transmission at small $\omega$ is
\begin{equation}
\label{T-small_omega_0<s<1}
\mathcal T(\omega\to0^+) = 
\begin{cases}
{\ \displaystyle\frac{\pi^2K^2\ (\omega/\omegac)^{2s}}{\Gamma^2(s)(K-K_*)^2}}&\text{for $K\ne K_*\, ,$}\\
\ \sin^2(\pi s)&\text{for $K=K_*\, .$}
\end{cases}
\end{equation} 
Hence, the transmission at the band edge has a discontinuity at $K=K_*$; it vanishes at the band edge for all $K\neq K_*$, but is finite for $K=K_*$. Indeed when $s=1/2$ and $K=K_*$, the transmission at the band edge is perfect; $\mathcal T(\omega \to 0^+) =1$. While the transmission is only a discontinuous function of $K$ at $\omega\to 0^+$, it changes rapidly close to the transition for any $\omega\ll\omegad$ (as seen in Fig.~\ref{Fig-T_s0p5}). 

Furthermore, for $1/2\le s<1$ the transmission function has a peak with perfect transmission (i.e.~transmission equal to one at a certain value of $\omega$) for all $K<K_*$, while no such perfect transmission peak exists for $K>K_*$. 
This is clearly visible in Fig.~\ref{Fig-T_s0p5}, where $\mathcal T(\omega)$ exhibits a large peak (going up to transmission of one) for all $K<K_*$ but is a broad function which never gets close to a transmission of one for $K>K_*$. 

For $0<s<1/2$, the situation is basically the same, except that there is a very small regime at $K>K_*$ with a double peak in the transmission.\cite{in-preparation}

\begin{figure}
\includegraphics[width=\linewidth]{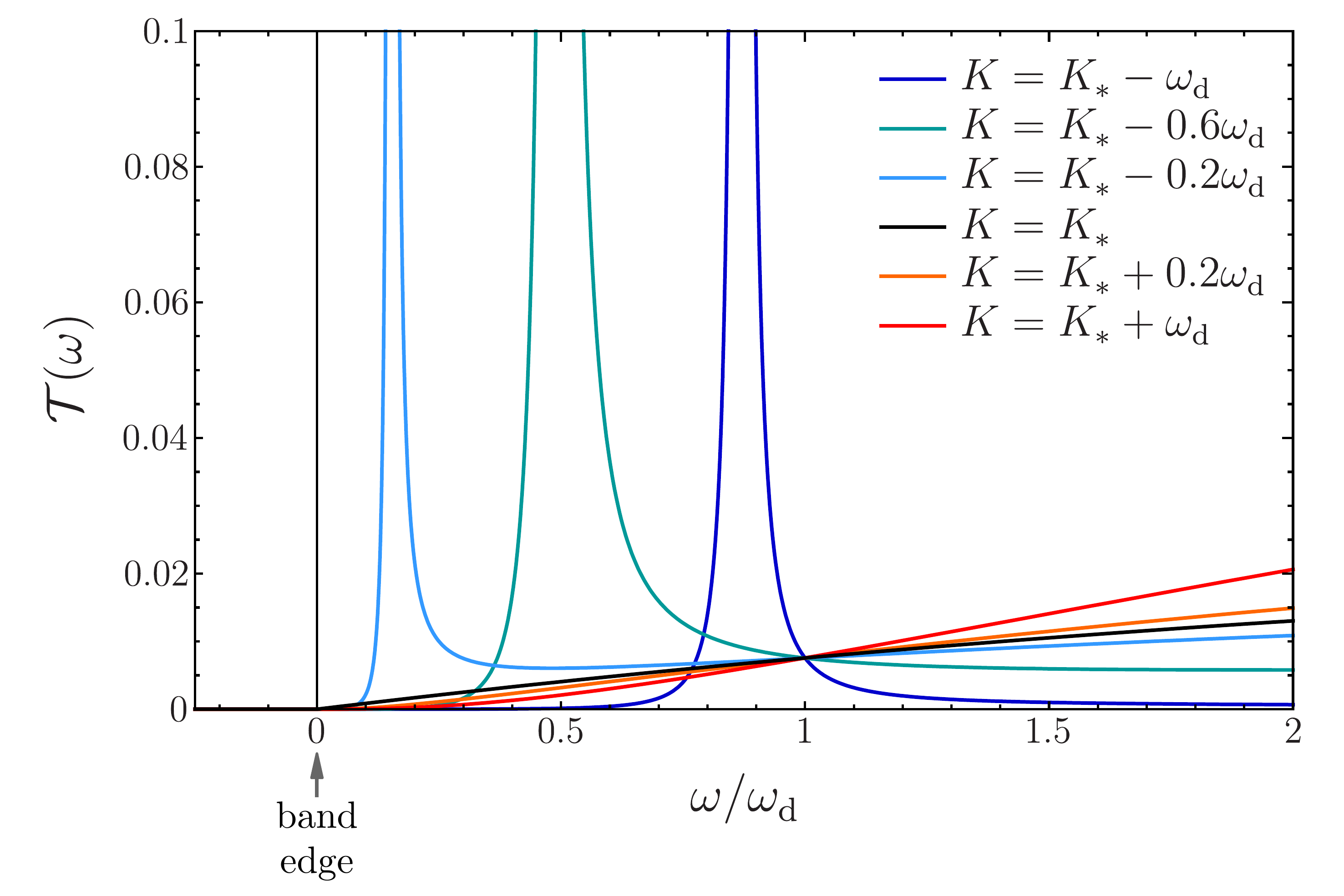}
\caption{Typical behaviour of the transmission as a function of energy for various values of the coupling, when $J(\omega)$ is given by Eq.~\eqref{J} with spectral exponent $s\ge1$. The parameters are $s=3/2$ and $\omegac=10\omegad$, but the plots are qualitatively similar for $s=1$ or other $s>1$. As the coupling approaches the critical value from below, the peak in the transmission moves towards the band edge, becoming narrower and narrower. For all $K< K_*$, the transmission features a peak whose maximum is at 1 (maximum not shown). When $K$ passes though the critical value $K_*$, this peak disappears.}
\label{Fig-T_s1p5}
\end{figure}

\subsection{Transmission for $s\ge1$}

For the spectrum in Eq.~\eqref{J} with $s\ge1$, the transmission at small $\omega$, has the same form as in Eq.~\eqref{T-small_omega_0<s<1} for $K \neq K_*$, but has a different functional form for $K=K_*$. This form at $K=K_*$ depends on whether $s=1$ or $s>1$. For $s=1$, it is 
\begin{equation}
\mathcal T(\omega\to0^+;K=K_*) = 
\left(\frac{\pi}{\ln(\omega/\omegac)}\right)^2
\end{equation} 
For $s>1$, it is
\begin{equation}
\mathcal T(\omega\to0^+;K=K_*) = 
\frac{\pi^2 \omegad^2 \,(\omega/\omegac)^{2(s-1)}}{(\omegad\Gamma(s-1) +\omegac\Gamma(s))^2}
\end{equation} 
Thus, for both $s=1$ and $s>1$ there is a discontinuous behaviour at small $\omega$. Unlike for $0<s<1$, both functional forms vanish at $\omega\to 0^+$, however they have completely different $\omega$-dependences at $K=K_*$ than at other $K$. 

In this case, there is an additional type of discontinuity in $\mathcal T(\omega)$, which can be seen in Fig.~\ref{Fig-T_s1p5}, and is described as follows. As $K$ approaches $K_*$ from below, the transmission peak moves towards the band edge. However, unlike for $0<s<1$, the peak gets narrower as it approaches the band edge, so its width vanishes when it reaches the band edge at $K=K_*$. In other words, just below the transition (when $K-K_*$ is small and negative), the transmission goes from zero at $\omega=0$ up to a peak of perfect transmission (transmission of one) and then drops back down to a transmission close to zero in an energy window of order $|K-K_*|$. In contrast, this peak is completely absent for all $K>K_*$ (where the peak has become the bound state). Thus, here the discontinuity at $K=K_*$ is the disappearance of this sharp peak in the transmission at small $\omega$.

\begin{figure}
\includegraphics[width=\linewidth]{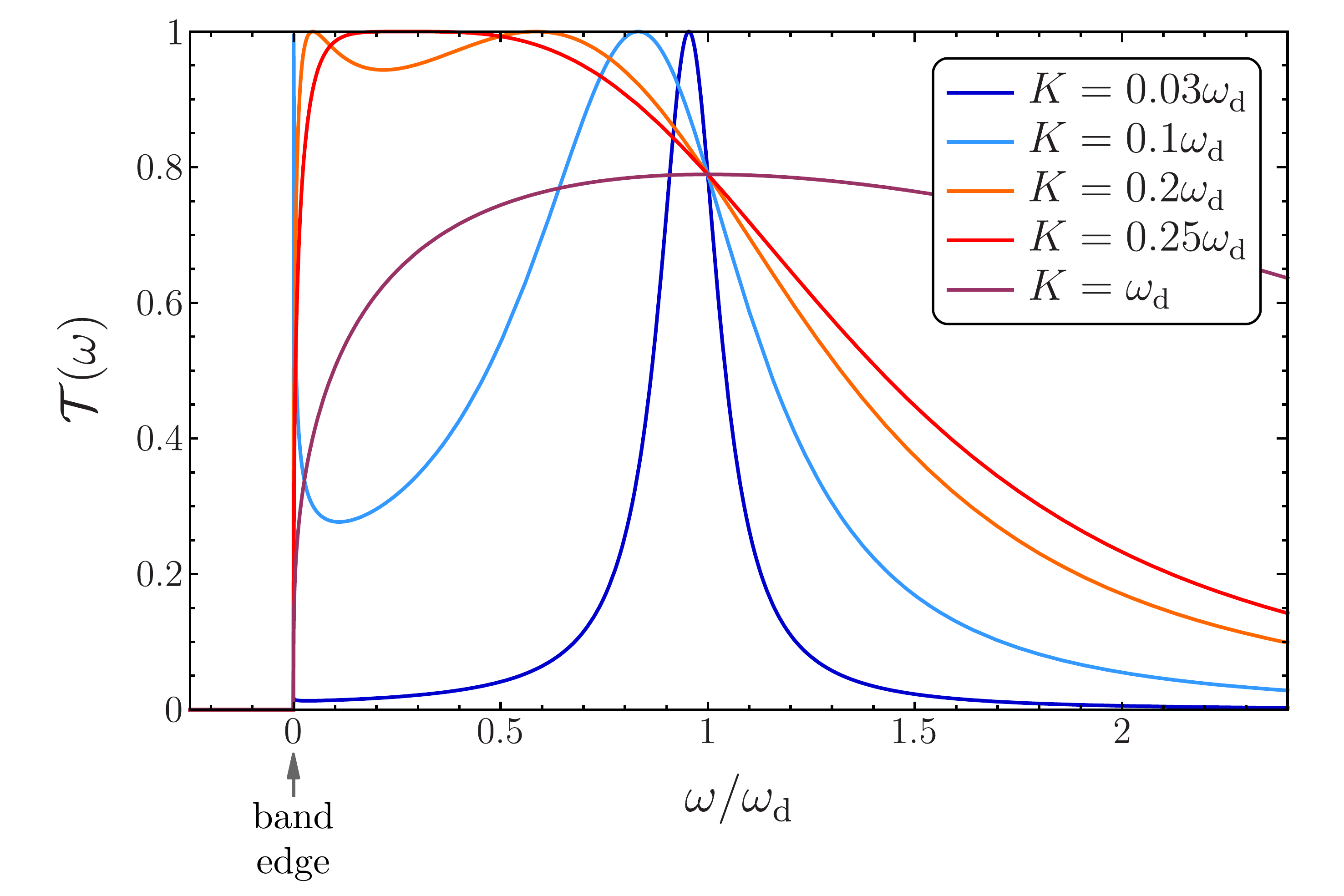}
\caption{Transmission as a function of energy for various values of the coupling, when $J(\omega)$ is given by Eq.~\eqref{J} with $s=0$ and $\omegac=10\omegad$. At small couplings, the transmission looks like a Lorentzian at the dot level, with an additional very narrow peak extremely close to the band edge (too narrow to be seen in the plot for $K=0.03\omegad$). As $K$ is increased, the two peaks broaden and move towards each other, before coalescing at $K\sim 0.25\omegad$.}
\label{Fig-T_s0}
\end{figure}

\begin{figure}
\includegraphics[width=\linewidth]{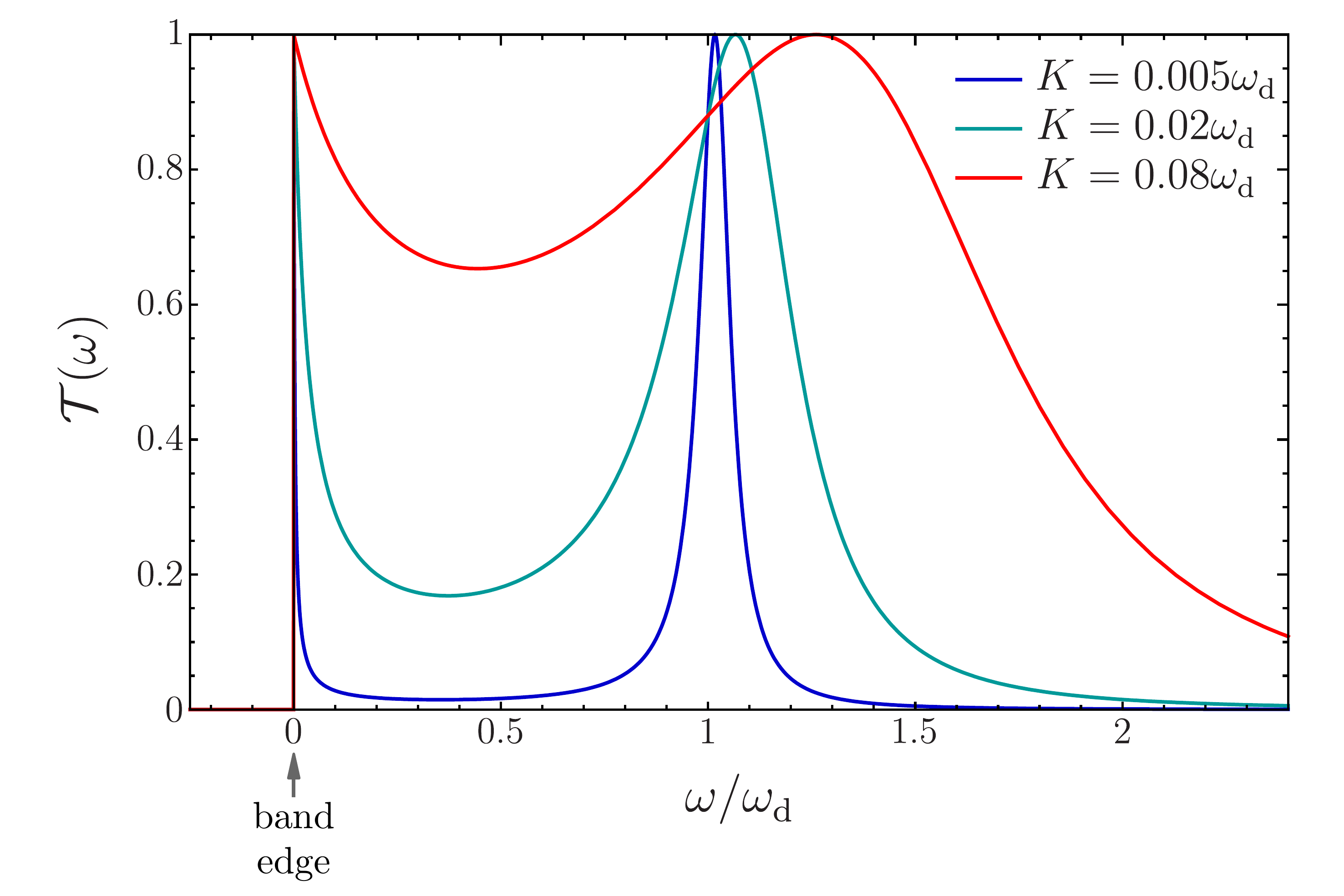}
\caption{Transmission as a function of energy for various values of the coupling, when $J(\omega)$ is given by Eq.~\eqref{J} with spectral exponent $-1<s<0$. The parameters are $s=-1/2$ and $\omegac=10\omegad$. In this case, the divergence of $J(\omega)$ at the band edge induces a peak of transmission at the band edge whose height does not depend on the value of the coupling. At small couplings, the transmission looks like a Lorentzian at the dot level, with an additional peak at the band edge.}
\label{Fig-T_s-0p5}
\end{figure}

\subsection{Transmission for $s\le0$}

The behaviour of the transmission function is very different when $J(\omega)$ does not vanish at the band edge, that is when $s\leq0$ for the spectral density in Eq.~\eqref{J}. The integrals become ill-defined if $s\le-1$, suggesting that $s\le -1$ is unphysical. However, the regime of $-1<s\le0$ is well-defined, with $s=-1/2$ being of particular relevance in the context of both one-dimensional reservoirs, and the quasi-particle spectrum in superconducting reservoirs.\cite{Shiba1973Jul,Basko2017}

For $-1<s\le0$, there is a bound state at \textit{all} values of the coupling. 
This was noted multiple times for $s=-1/2$, see e.g.~Refs.~[\onlinecite{Shiba1973Jul,Angelakis2004,Basko2017}], but is equally easily seen for other $s\leq0$ by looking at the condition for the bound state's existence, $\Omega(\omega\to0^-)>0$, just above our Eq.~\eqref{K_*-our-spectrum}. If the spectral function does not vanish at the band edge ($-1<s\leq0$ here), then this condition is always satisfied. Hence the critical coupling is zero for all $s\le0$, as shown in Fig. \ref{Fig-Transition}.

For $s=0$, the transmission is shown in Fig.~\ref{Fig-T_s0}. For all finite coupling $K$, it goes to zero at the band edge as follows
\begin{equation}
\mathcal T(\omega\to0^+)=\left(\frac{\pi}{\ln(\omega/\omegac)}\right)^2.
\end{equation} 
For $s<0$, the transmission is shown in Fig.~\ref{Fig-T_s-0p5}. The divergence of $J(\omega)$ at the band edge implies that the transmission function takes a finite value at this point,
\begin{equation}
\mathcal T(\omega\to0^+)\longrightarrow \sin^2(\pi s),
\end{equation}
for all finite $K$. This is seen in Fig.~\ref{Fig-T_s-0p5} where there is a peak at the band edge for all values of the coupling parameter. Thus, even for very small coupling, where one would guess that the transmission would be a narrow Lorentzian centred at the dot level, there is a second narrow peak at the band edge when $s$ is negative. 

\section{Interpretation as a Lamb shift}
\label{Sect:lamb}

To get a more intuitive feel for the physics of the model, this section provides a qualitative interpretation of the Lamb shift in terms of level repulsion between the dot level and the reservoir's continuum of states.

In the weak coupling limit, Fermi's golden rule tells us that the coupling to a continuum has two effects on the discrete levels of a quantum system.\cite{Cohen-TannoudjiGoldenRule} Firstly, the energy levels of the system are broadened to become resonances because these states acquire a finite lifetime. Secondly, the coupling shifts the energies of these levels; this is known as a Lamb shift. The usual weak-coupling formula for the Lamb shift of the quantum dot level is $\Lambda(\omegad)$, with $\Lambda(\omega)$ given in Eq.~\eqref{Lambda}. This fits with the exact result in Eq.~\eqref{Lambda} when the coupling is weak enough that the physics is dominated by $\omega\simeq\omegad$.

The Lamb shift can be interpreted at the handwaving level in terms of the level repulsion between the dot level and individual continuum levels. Level repulsion between the dot level and a continuum level with higher energy will shift the dot level down in energy (with the continuum level being shifted up slightly). At the same time, the repulsion between the dot level and a continuum level with lower energy will shift the dot level up in energy. The Lamb shift is the sum of all of these small shifts. If the continuum has a constant density of states (wide-band limit), then the shifts up and down cancel and there is no Lamb shift. However, if the continuum has a higher density of states above the dot level than below it (as in Figs.~\ref{Fig-T_s0p5} and~\ref{Fig-T_s1p5}), the Lamb shift is negative and moves the dot level to lower energies. 

\begin{figure*}
\includegraphics[width=\linewidth]{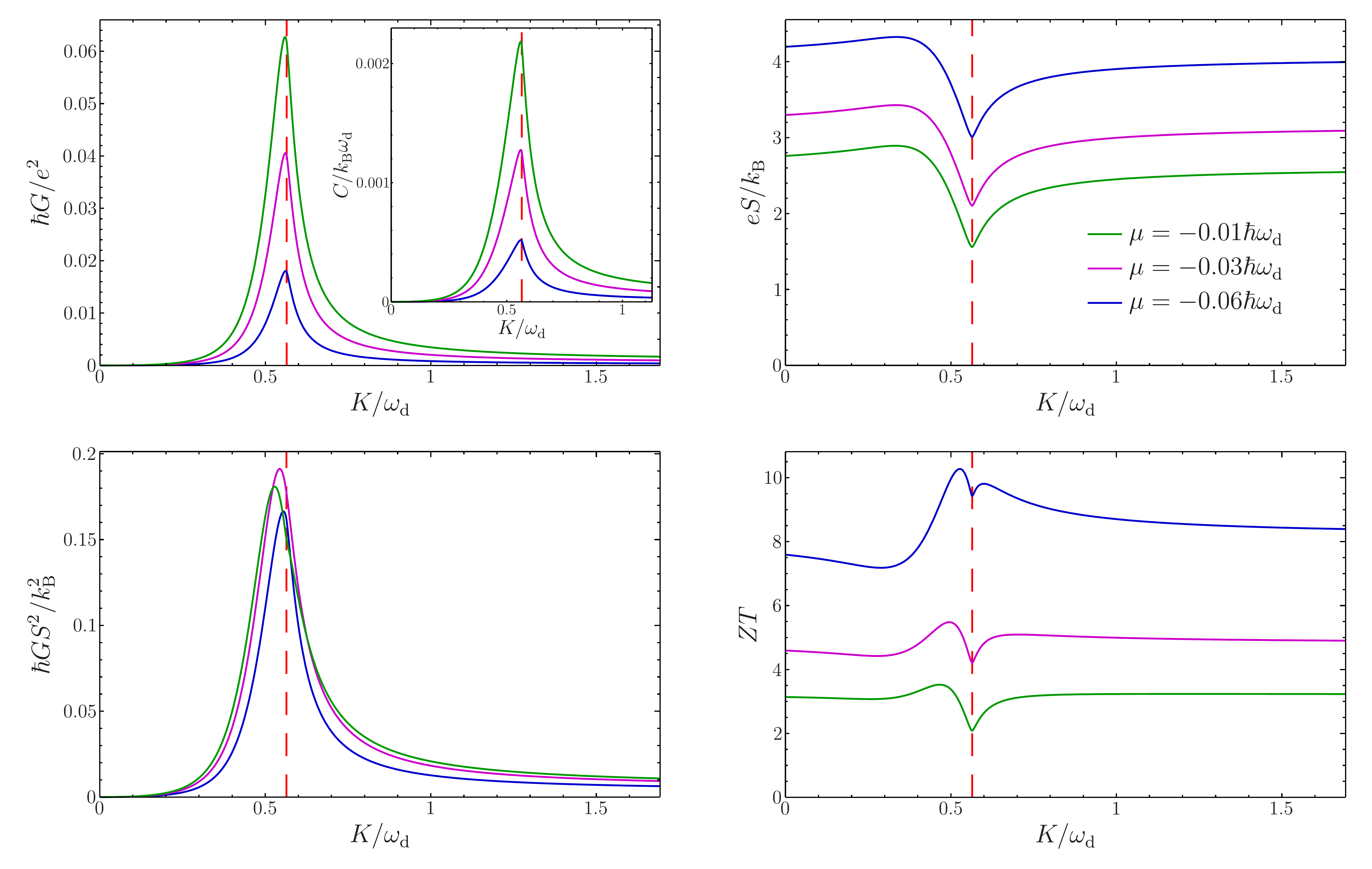}
\caption{Transport coefficients and thermoelectric parameters versus coupling for different values of the chemical potential. The parameters are $s=1/2$, $\omegac=10\omegad$ and $T=0.03\omegad/k_\mathrm B$. The transport coefficients are plotted in dimensionless units, so the electrical conductance is shown as $\hbar G\big/e^2$, the heat conductance as $C\big/(k_\mathrm B\omegad)$, and the Seebeck coefficient as $eS/k_\mathrm B$. The thermoelectric power factor in dimensionless units is $hGS^2\big/k_\mathrm B^2$. The dashed red vertical line indicates the critical coupling $K_*$, with $K_*/\omegad\simeq 0.56$ for these parameters.}
\label{Fig-Transport}
\end{figure*}

The transmission given in Eq.~\eqref{transmission} exhibits a perfect peak ($\mathcal T(\omega)=1$) for $\omega$ in the continuum such that $\omega=\omegad-\Lambda(\omega)$. For $s>0$, as in Fig.~\ref{Fig-T_s0p5}, this peak is clearly visible for $K<K_*$ and it moves to lower energies as the coupling is increased. This corresponds to an increasingly negative Lamb shift of the dot level, to which the coupling also gives a finite lifetime, broadening it into a resonance. When the coupling reaches the critical value $K_*$, the peak sits exactly at the band edge, so at this and only this coupling, the transmission at the band edge can be finite. When the coupling $K$ becomes larger than $K_*$ the Lamb shift is so large that it has moved the peak out of the band. One can naively interpret this as the Lamb shift having pushed the dot level out of the band, at which point the level becomes a bound state with energy $\omega_*$ such that $\Omega(\omega_*)=0$, see Eq.~\eqref{Omega}. However, more precisely, the dot state is then a superposition of the bound state and continuum states, so there is still transmission through the dot at $K>K_*$, but it no longer exhibits a peak with transmission equal to one. 

This handwaving argument does not work very well for $s<1/2$. For $0<s<1/2$, it does not capture the very small region for $K>K_*$ where there are two peaks in the transmission.\cite{in-preparation} Furthermore it fails completely for $s<0$, the usual weak-coupling arguments\cite{Cohen-TannoudjiGoldenRule} do not reproduce the bound state that is always present in the exact solution. However, the handwaving description above at least gives an indication of why the transmission peak at $\omega=\omegad$ in Fig.~\ref{Fig-T_s-0p5} moves to higher energies as the coupling is increased. This is because the Lamb shift is positive when the density of states is larger below the dot level than above it. At the same time, the divergence of $J(\omega)$ at the band edge induces another peak at this point. Indeed, the divergences in the numerator and the denominator of the transmission function in Eq.~\eqref{transmission} cancel each other, resulting in $\mathcal T(\omega\to0^+)$ taking a finite value irrespective of the coupling parameter $K$.

\section{Thermoelectric transport}
\label{Sect:thermoelectric}

When differences in temperature and electrochemical potential are small compared to the average temperature, linear response theory tells us that all transport properties are given by\cite{ReviewBCSW} the electrical conductivity $G$, thermal conductivity $C$, and Seebeck coefficient $S$. The quality of a thermoelectric device is then determined by two quantities --- the thermoelectric power factor $GS^2$ and the dimensionless figure of merit $ZT=GS^2T/C$, where $T$ is the average temperature. Larger $GS^2$ and larger $ZT$ implies a better thermoelectric device.

\begin{figure}[t]
\includegraphics[width=\linewidth]{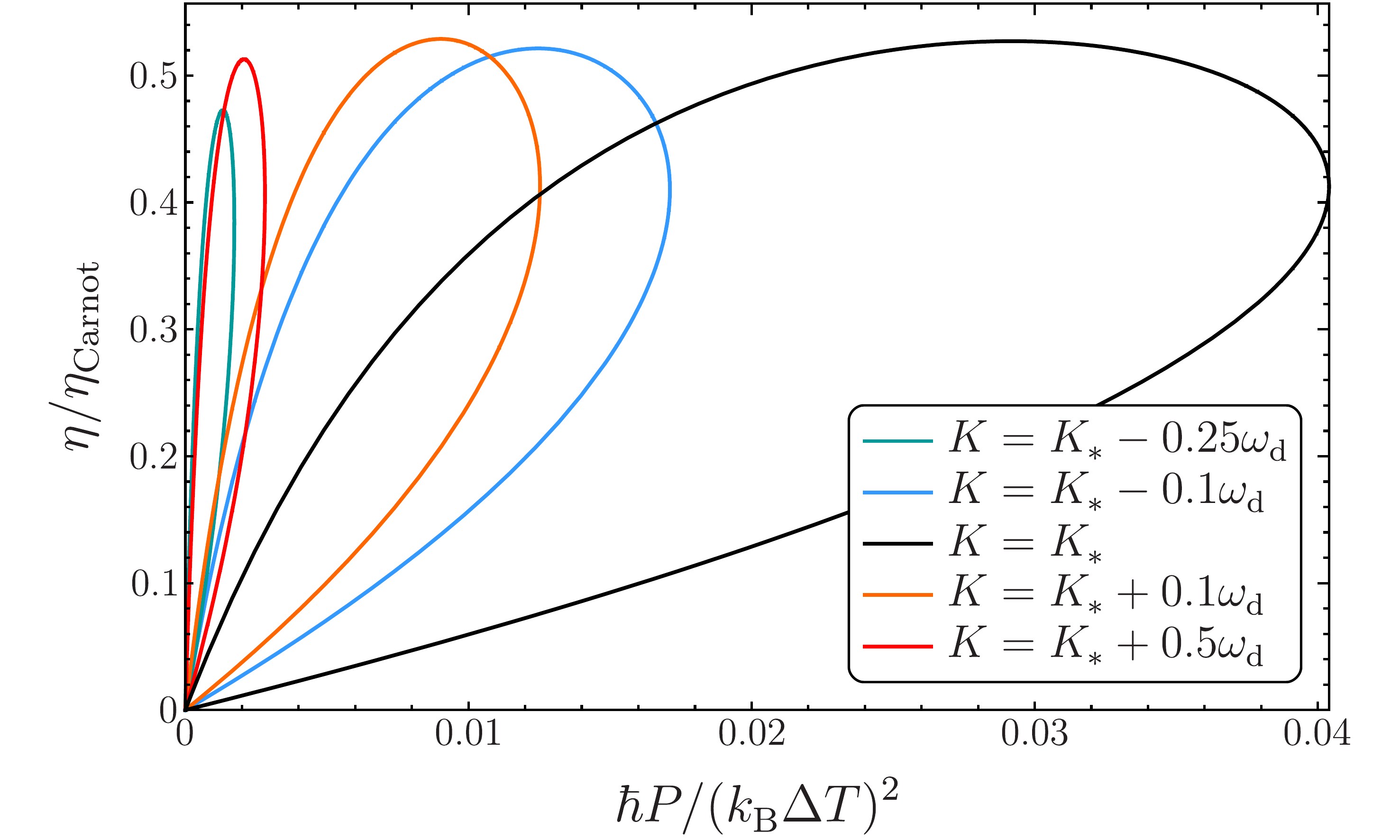}
\caption{
The plot of efficiency versus power output given by Eq.~\eqref{Eq:eta-vs-P} when $P_\mathrm{max}$ and $ZT$ are found from the Fano-Anderson model with dot-reservoir coupling $K$. The parameters are $s=1/2$, $\omegac=10\omegad$, $\mu=-0.06\omegad$ and $T=0.03\omegad/k_\mathrm B$. The curves have the typical ``loop'' form, and one clearly sees that the power output is much bigger when $K=K_*$, without a significant change of efficiency.}
\label{Fig-etaxP}
\end{figure}

The power factor $GS^2$ determines the maximum power output in the linear response regime,\cite{ReviewBCSW}
\begin{equation}
P_\mathrm{max}=\tfrac14\,GS^2 \,\Delta T^2 
\end{equation}
where $\Delta T$ is the temperature difference across the thermoelectric device. This is determined by defining the power output as $P=j_\mathrm{steady}^{(\mathrm N)}\Delta \mu$, for potential difference $\Delta \mu=\mu_\mathrm R-\mu_\mathrm L$, and tuning $\Delta \mu$ to maximize $P$. The dimensionless figure of merit $ZT$ is a measure of a thermoelectric's maximum efficiency. For any given thermoelectric device, the efficiency (defined as power output over heat input) as a function of power output is\cite{ReviewBCSW}
\begin{equation}
\eta(P) =\frac{\eta_\mathrm{Carnot}}2\ \frac{P/P_\mathrm{max}}{1+2/ZT\mp\sqrt{1-P/P_\mathrm{max}}}
\label{Eq:eta-vs-P}
\end{equation}
where $\eta_\mathrm{Carnot}$ is the Carnot efficiency for this $\Delta T$. This function takes the well-known form of a ``loop'' and is plotted in Fig.~\ref{Fig-etaxP} for various values of the dot-reservoir coupling.

Hence, for any given $P/P_\mathrm{max}$ the efficiency is increased by increasing $ZT$. There are two well-known results of this formula. Firstly, the efficiency at maximum power is $\eta(P_\mathrm{max}) = \eta_\mathrm{Carnot}\, ZT \big/ (2ZT +4)$. Secondly, the maximum efficiency --- achieved by tuning $P$ to maximize $\eta$ (physically this is done by tuning $\Delta\mu$) --- is
\begin{equation}
\eta_\mathrm{max}=\eta_\mathrm{Carnot} \big(\sqrt{ZT+1} -1\big)\big/\big(\sqrt{ZT+1} +1\big),
\end{equation}
so one requires $ZT \to \infty$ to achieve Carnot efficiency. The power output at $\eta_\mathrm{max}$ is
\begin{equation}
P=2P_\mathrm{max}\,\sqrt{ZT+1}\big/\big(1+ \tfrac12ZT+\sqrt{ZT+1}\big). 
\label{Eq:P-at-etamax}
\end{equation}
Thus, if $GS^2$ increases (increasing $P_\mathrm{max}$) without much reduction of $ZT$, the power output improves without much loss of efficiency.

Taking the linear-response regime of Eqs.~(\ref{j^(N)_dc},\ref{j^(E)_dc}), we find that the electric conductance, $G=e^2I_0$, the thermal conductance, $C=(I_2-I_1^2/I_0)\big/T$, and the Seebeck coefficient, $S=I_1\big/(eTI_0)$, where $e$ is the electron charge, $T$ is the average temperature of the two reservoirs, and 
\begin{equation}
I_n=\int_{-\infty}^\infty\frac{\mathrm d\omega}{2\pi}\,(\omega-\mu)^n\ \mathcal T(\omega)(-f'(\omega)),
\label{I-integrals}
\end{equation}
with $f'(\omega)$ being the derivative of the Fermi distribution; $-f'(\omega)=\beta/(2\cosh(\beta(\omega-\mu)/2))^2$. The two thermoelectric parameters (power factor and figure of merit) then read $GS^2 = I_1^2\big/(T^2 I_0)$, and $ZT=I_1^2\big/(I_0I_2-I_1^2)$.

From these equations, we see that the change of behaviour of the transmission at low energies leads to rapid changes for the transport coefficients when one considers the electrochemical potential below the band edge ($\mu <0$). Indeed, $-f'(\omega)$ is basically an energy filter centered on $\mu$ of width $\sim T$ that is superimposed on the transmission function. If one then takes $\mu$ to be negative and $T$ of the order of $|\mu|$ (or smaller), the transport coefficients will be dominated by the small $\omega$ behaviour of $\mathcal T(\omega)$. 
These coefficients carry a signature of the discontinuities of ${\cal T}(\omega)$ in their rapid change with $K$ when it is close to $K_*$, see Fig.~\ref{Fig-Transport}. In some cases the change is so rapid, that the curves looks discontinuous, however they only becomes strictly discontinuous in the $\mu\to 0$ and $T\to 0$ limit, which is also the limit where $C$ and $S$ become vanishingly small.

The conductances ($G,C$) and thermoelectric power factor ($GS^2$) exhibit huge peaks at $K\simeq K_*$, whereas the Seebeck coefficient ($S$) and the figure of merit ($ZT$) have small dips at this point. The peaks in $G$, $C$ and $GS^2$ are the result of peaks in the $I_n$-functions very close to $K\to K_*$ for the case we consider, with the peaks having a similar magnitude for all $n$. To see the origin of these dips in $S$ and $ZT$, we note that both quantities consist of ratios of $I_n$-functions; see the formulas for $S$ above Eq.~\eqref{I-integrals} and $ZT$ below Eq.~\eqref{I-integrals}. Hence for $K \simeq K_*$, $S$ and $ZT$ are each a ratio in which there is competition between a peak in the numerator and a peak in the denominator, where both peaks have similar magnitude. In all cases, we have studied it is the peak in the denominator that is a little stronger, so there is a small dip in $S$ and $ZT$. However, we have not found an argument for why this is so, thus we cannot rule out a small peak in other parameter regimes. In the limit $\mu\to 0$ and $T\to 0$, the peaks and dips become discontinuities in the derivative of the function in question, and sit exactly at $K_*$.

From the point of view of engineering the thermoelectric response, one see that the thermoelectric power factor $GS^2$ has a huge peak near at $K\simeq K_*$, while $ZT$ is not very strongly varying. Hence, if one has a thermoelectric system of this type with the electrochemical potential close to or below the band edge, one can get much bigger power for approximately the same efficiency by taking $K\simeq K_*$, see Fig.~\ref{Fig-etaxP}. 

\section{Conclusions}
\label{Sect:conclusions}

We consider a quantum dot coupled to reservoirs with band gaps. For situations where an infinite-lifetime bound state appears in the band gap when the dot-reservoir coupling exceeds a critical value (i.e.~when the reservoir spectra that vanish at the band edge), we show that this transition induces discontinuities in the dot's transmission. 
These are a sign of changes in the continuum state's properties when the bound state emerges from the continuum. 
This has a strong signature in the electric and thermoelectric transport properties, whenever the reservoirs' electrochemical potentials are close to or below the band edge. Under such conditions the system's optimal thermoelectric response is close to the critical dot-reservoir coupling, $K\simeq K_*$; with a huge increase in power output accompanied by a small change in efficiency. 

For reservoir spectra that diverge at the band edge, it is known that there is a bound state for all coupling. We show that the dot's transmission has a peak at the band edge, even at arbitrarily weak coupling. This peak will dominate transport whenever the electrochemical potential is near the band edge. The usual argument, that the dot's transmission is a Lorentzian centred at the dot level, will give erroneous transport properties in such situations. 

The richness of this physics could not be guessed from the usual weak-coupling arguments, which suggests that other surprises may await us in the strong-coupling limit when we add electron-electron interactions.

\begin{acknowledgments}
We thank D. Basko, M. Holzmann, N. Lo Gullo, and K. Saito for stimulating discussions and useful comments. We thank the anonymous referees for drawing our attention to a number of works, particularly Ref.~[\onlinecite{Maciejko2006Aug}] and various works of Zhang and co-workers. 
This work was supported by the grant ANR-15-IDEX-02 via the Universit\'e Grenoble Alpes QuEnG project.
\end{acknowledgments}

\appendix

\section{Calculation of $\phi(t)$}
\label{Sect:inverse-laplace}

We define $\phi(t)$ as the inverse Laplace transform of the prefactor in Eq.~\eqref{Laplace_solution_D(z)}; $1/(z+\ii(\omegad+\Sigma(z)))$. This function appears in nearly all observables, and contains the physics of the bound states. 
It was calculated in Ref.~[\onlinecite{Zhang2012}], where it was denoted $u(t-t_0)$. 
However, we briefly review it here to fix notation, and because this inverse Laplace transform gives access to the full time-dependent solution of the problem.

The inverse Laplace transform is given by an integration\cite{BookAppel} as in Fig. \ref{Fig-Contour}b. The solution then has two parts: the first is the integral along the branch cuts (the contribution of the continuum of states), and the second comes from poles (bound states). The branch cut is always present, while the poles are only there if $z+\ii(\omegad+\Sigma(z))=0$ has a solution. Such a solution is purely imaginary and so is given by the (real) zeros of the function $\Omega(\omega)$ in Eq.~\eqref{Omega}. The zeros of $\Omega(\omega)$ only occur at values of $\omega$ where $J(\omega)=0$, otherwise the integral in Eq.~\eqref{Omega} is divergent, so poles only occur at energies in the band gaps of the reservoirs.

One eventually finds that
\begin{equation}
\phi(t)=\integ{\mathrm B}{}\omega S(\omega)\ee^{-\ii\omega t}+\sum_n Z_{*n}\ee^{-\ii\omega_{*n} t},
\label{phi(t)}
\end{equation}
where the integral is taken over the continuum (branch cuts) and the sum is over all bound states (poles). The continuum contribution contains $S(\omega)$ which one can show\cite{Xiong2015Aug} is the dot's local density of states,
\begin{equation}
S(\omega)=\frac{J(\omega)}{(\omega-\omegad-\Lambda(\omega))^2+\pi^2J(\omega)^2},
\label{S}
\end{equation}
where $\Lambda(\omega)$ corresponds to the Lamb shift which accounts for the renormalization of the dot level due to the coupling to the reservoir. It is the real part of $\Sigma(\varepsilon-\ii\omega)$ in Eq.~\eqref{Sigma}, and it is given by $\Lambda(\omega)=\Lambda_\mathrm L(\omega)+\Lambda_\mathrm R(\omega)$, with the Cauchy principal value integral
\begin{equation}
\Lambda_\alpha(\omega)=P\hskip-3.75mm\integ{}{}\omega'\,\frac{J_\alpha(\omega')}{\omega-\omega'}.
\label{Lambda}
\end{equation}
The bound state contribution in Eq.~\eqref{phi(t)}, contains $Z_{*n}$ which is the dot level's overlap with the $n$th bound state, whose energy is $\omega_{*n}$. It reads
\begin{equation}
Z_{*n}=
\begin{cases}
\ \displaystyle\left(1+\integ{\mathrm B}{}\omega\frac{J(\omega)}{(\omega-\omega_{*n})^2}\right)^{-1}&\text{for $K>K_{*n},$}\\
\ \ \ 0&\text{for $K<K_{*n}.$}
\end{cases}
\label{Z_*-definition}
\end{equation}
where $K_{*n}$ is the critical coupling above which the $n$th bound state appears. For the spectral density in Eq.~\eqref{J} which can only give rise to one bound state, the precise nature of the discontinuity at $K=K_*$ is seen by taking $K-K_*$ to be small.
Then for $0<s<1$, 
\begin{equation}
Z_*=
\begin{cases}
\displaystyle\frac\omegac s \ \big[B(s)\big]^{1/s}\,
\big[\Gamma(s) \,(K-K_*)\big]^{(1-s)/s} \!\!&\text{for $K>K_*$\,,}\\
\ \ 0&\text{for $K<K_*$\,,}
\end{cases}
\label{Z_*-small-K-K*}
\end{equation}
where $B(s)= \sin(\pi s)\big/(\pi K)$,
while for $s>1$,
\begin{equation}
Z_*=
\begin{cases}
(s-1)\big/(s-1+\omegad/\omegac) \!\!&\text{for $K>K_*$\,,}\\
\ \ 0&\text{for $K<K_*$\,.}
\end{cases}
\label{Z_*-small-K-K*}
\end{equation}

\section{Dot dynamics}
\label{Sect:dot-occupation}

The dynamics of the dot occupation $n(t)$ is not the subject of this article, they have been extensively discussed elsewhere (see the works cited in the introduction). However, we need $n(t)$ to use in the continuity equation, which allows us to greatly simplify the results for the currents that interest us.

Taking the initial state in Eq.~\eqref{Eq:product-state}, the average number of electrons on the dot is
\begin{equation}
n(t)=\langle\hat d^\dagger(t)\hat d(t)\rangle,
\end{equation}
where $\langle \cdots \rangle = \mathrm{Tr}\big[ \cdots \ \hat\rho (t=0)\big]$. The exact dynamics of the field operators derived above provide a full solution of the model. So the occupation for any time is
\begin{equation}
n(t)=n_0|\phi(t)|^2+\integ{\mathrm B}{}\omega J(\omega)F(\omega)|\psi(t,\omega)|^2,
\label{n(t)-formal}
\end{equation} 
where $n_0$ is the initial occupation of the dot.

\begin{figure}
\includegraphics[width=\linewidth]{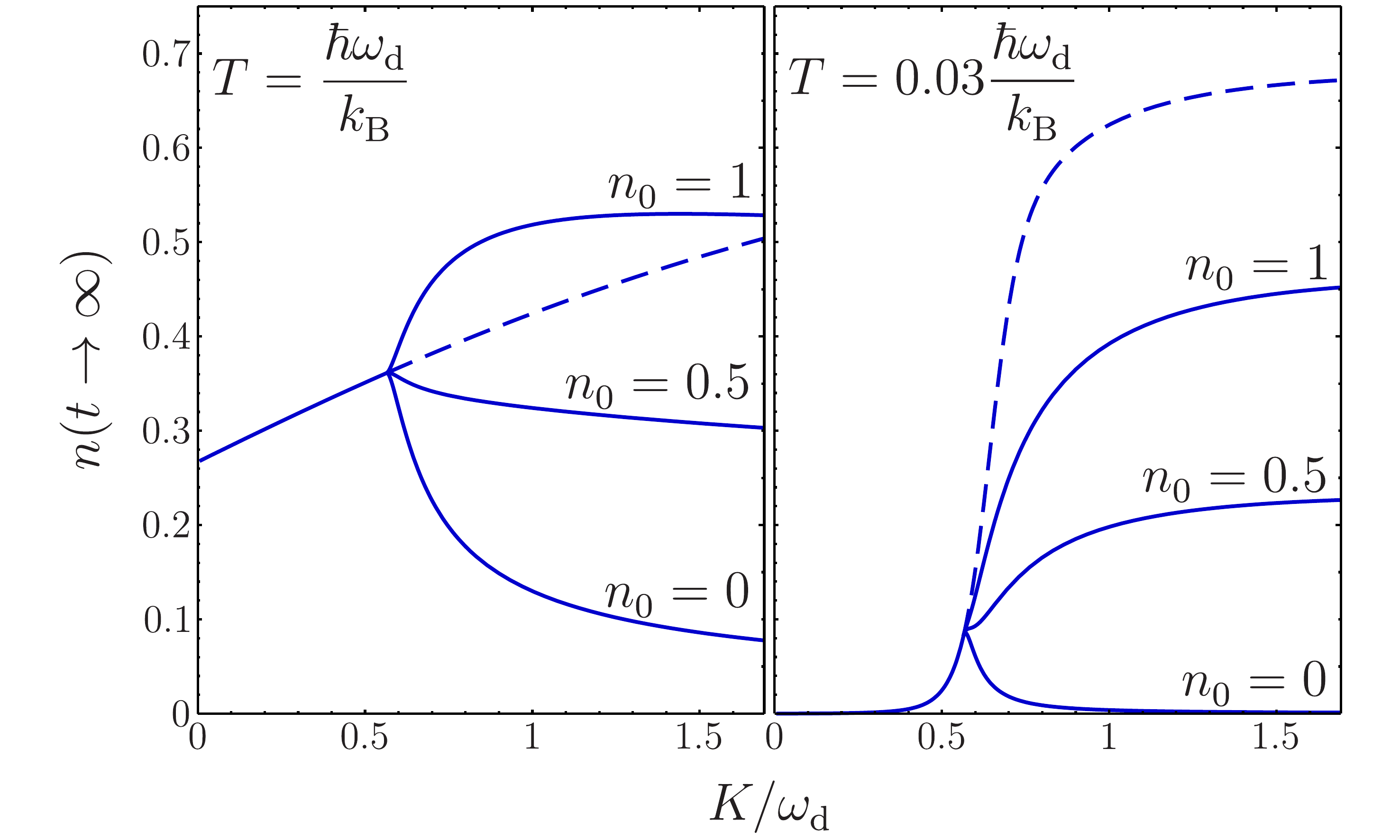}
\caption{Long-time dot occupation versus the system-reservoir coupling $K$, when the two reservoirs have the same temperature and bias. The spectral density for this plot is given by Eq.~\eqref{J} with $s=1/2$, $\omegac=10\omegad$, $\mu=-0.01\omegad$. The solid lines representing the occupation a long-time after a quench, see Eq.~\eqref{n_long-time}, and exhibit a clear change of behaviour for $K=K_*\simeq0.564\omegad$. When $K>K_*$, the occupation at arbitrarily long times depends on the dot's initial occupation $n_0$. The dashed lines are the occupation a long time after an adiabatic preparation, see Eq.~\eqref{nsteadyeq}, and do not show the transition.}
\label{Fig-nsteady}
\end{figure}

In section~\ref{Sect:Long-time-currents}, we used the time-derivative of this result in the continuity equation to derive Eq.~\eqref{A-result}, which greatly simplifies formulae for long-time currents. We note in passing that it also greatly simplifies the formula for the long-time dot occupation, which then reads
\begin{multline}
n (t\to \infty)=\integ{\mathrm B}{}\omega S(\omega)F(\omega)
+\sum_{n,m} \, M_{nm} \cos (\omega_{nm}t).
\label{n_long-time}
\end{multline}
Here the integral is over the the bands, and $M_{nm}$ depends on the initial dot occupation $n_0$, see Eq.~\eqref{M_nm}. If there are no bound states, $n(t\to\infty)$ is independent of the initial dot occupation $n_0$. When a bound state exists, the long-time occupation $n(t\to\infty)$ depends on $n_0$ (as for $K>K_*$ in Fig.~\ref{Fig-nsteady}), because the dot gets partially trapped in its initial state forever. If there are two or more bound states (which is \textit{not} the case for the spectrum in Eq.~\eqref{J}), then the dot occupation is known to oscillate forever\cite{Xiong2015Aug,Yang2015Oct,Ali2017Mar} at frequencies given by the energy difference between the bound states. Since electrons are conserved, the currents also exhibit oscillations at these frequencies for all times as seen in Eq.~\eqref{j^(N)_steady}. 

\section{Adiabatic preparation and comparison with Keldysh results}
\label{Sect:adiabatic}

Many observables are known to depend on the initial preparation of the quantum dot.\cite{Yang2015Oct} By comparing the initial state in Eq.~\eqref{Eq:product-state} with an adiabatic preparation\cite{Dhar2006Feb,Ali2015Dec,Ali2017Mar} of the dot-lead system, we will see that the steady-state DC currents are the same, suggesting that DC currents are insensitive to the initial preparation.

In experiments on quantum dots or molecular nano-structures it may be difficult to initially turn-on the dot-reservoir coupling rapidly enough to treat it as an initial quench that justifies Eq.~\eqref{Eq:product-state}. At the same time, weak inelastic scattering effects (phonons, etc), 
whose description is beyond Eq.~(\ref {H}), will give the bound state a long but finite lifetime. Hence, one might consider the initial state is \textit{adiabatically prepared}; i.e.~even the bound states have relaxed to their steady state\cite{equilib-note} at the start of the experiment. In reality, most experiments will fall somewhere between the initial quench assumed in Eq.~\eqref{Eq:product-state} and this adiabatic preparation.

By comparing adiabatic preparation\cite{Ali2015Dec,Ali2017Mar} to the case of an initial quench, we show that the steady-state DC currents do not depend on the type of preparation, even if many other observables do.\cite{Yang2015Oct} We also clarify why Eq.~\eqref{j^(N)_steady} differs from the Meir-Wingreen formula\cite{Meir1992} derived from Keldysh theory. A third type of preparation\cite{Cini1980Dec,Stefanucci2007May} called ``partition free'' is not addressed here.

Similarly to Ref.~[\onlinecite{Dhar2006Feb}], we model the very slow equilibration of the bound state by adding an infinitesimally weak coupling $\eta\,\kappa_\alpha$ between the dot state and the reservoir $\alpha$ at all energies. Here $\eta$ is taken to be a very small coupling constant, while $\kappa_\alpha$ (which is of order one) determines the relative strength of this coupling to the different reservoirs. This is a crude way of mimicking the weak inelastic effects, but will be sufficient for our purposes. For this, we replace $J_\alpha(\omega)$ by 
\begin{equation}
\tilde J_\alpha(\omega)=J_\alpha(\omega)+\eta\kappa_\alpha
\end{equation}
in all formulae (indicating them with a tilde). As $\tilde J(\omega)$ extends over all energies, all formulae derived in this work apply if one extends the integrals to all $\omega$, and drops the $Z_*$-terms. Then, bound states are replaced by narrow Lorentzian resonances in the local density of states, which become delta peaks in the limit $\eta\to0$.

To model the adiabatic preparation, we assume the system obeys Eq.~\eqref{Eq:product-state} at $t=0$, but that we start the experiment at time $t$ that is so much latter that the system (including any bound states) has arrived at the steady-state. For this, we take $t\to\infty$, and then $\eta\to0$ to recover the bound state's effect (including the $Z_{*n}$ terms) on evolution between the experiment's start ($t$) and its end ($t+\tau$), while ensuring that bound states had relaxed at the experiment's start.

Then the self-energy is $\tilde\Sigma(x-\ii\omega)=\Sigma(x-\ii\omega)-\ii\pi\eta(\kappa_\mathrm L + \kappa_\mathrm R)\sgn(x).$
The Lamb shift is unchanged as the additional term in the self-energy is purely imaginary. The steady-state occupation for finite $\eta$ reads $\tilde n (t\to \infty)=\integ{\mathrm B}{}\omega\tilde S(\omega)\tilde F(\omega)$. Taking the limit $\eta\to0$ for those $\omega$ where $J(\omega)=0$ gives\cite{Dhar2006Feb}
\begin{align}
&\lim_{\eta\to0}\tilde F(\omega)=
\frac{\kappa_\mathrm Lf_\mathrm L(\omega)+\kappa_\mathrm Rf_\mathrm R(\omega) }{\kappa_\mathrm L+\kappa_\mathrm R},\\
&\lim_{\eta\to0}\tilde S(\omega) 
=\sum_n Z_{*n} \, \delta(\omega-\omega_{*n}).
\end{align}
This yields
\begin{equation}
n (t\to \infty)=\integ{\mathrm B}{}\omega S(\omega)F(\omega)+\sum_n Z_{*n}F(\omega_{*n}),
\label{nsteadyeq}
\end{equation}
where the integral is taken over the bands. This is a continuous function of dot-reservoir coupling (the dashed line in Fig.~\ref{Fig-nsteady}). Unlike for the initial quench in Eq.~\eqref{n_long-time}, there are never oscillating terms here. Similarly, there are no oscillatory terms in the currents, and so 
\begin{equation}
j_\alpha^{(\mathrm N)}(t\to \infty)=j_\mathrm{DC}^{(\mathrm N)}\ \ \text{ and }\ \ j_\alpha^{(\mathrm E)}(t\to \infty)=j_\mathrm{DC}^{(\mathrm E)},
\end{equation}
where $j_\mathrm{DC}^{(\mathrm N)}$ and $j_\mathrm{DC}^{(\mathrm E)}$ are given by Eqs.~(\ref{j^(N)_dc},\ref{j^(E)_dc}). This result coincides with the well-known Keldysh result of Meir and Wingreen\cite{Meir1992} in the non-interacting limit. Indeed, most works that we know of that use the Keldysh technique consider adiabatic preparation, and so do not show oscillations in the dot occupation or the current.

While there is no discontinuity in $n(t)$ when bound states emerge from the continuum, the dot does acquire infinite-time correlations, which can be seen in the correlation function\cite{Ali2015Dec,Ali2017Mar} $G(\tau)=\lim_{t\to\infty}\big[ \langle\hat n(t+\tau)\hat n(t)\rangle-\langle\hat n(t+\tau)\rangle\langle\hat n(t)\rangle\big]$. Taking $t\to\infty$, then $\eta\to0$ and finally large $\tau$, one finds
\begin{equation}
G (\tau\to \infty)= \sum_{n,m}Z_{*n} Z_{*m}F_{*n}\left(1-F_{*m}\right)\ee^{\ii(\omega_{*n}-\omega_{*m})\tau},
\label{G_steady}
\end{equation}
where we write $F(\omega_{*n})$ as $F_{*n}$ for compactness. This is zero in the absence of bound states, it is constant if there is one bound state, and it oscillates if there are multiple bound states. We can expect similar correlations in the current at large time differences, however these are unlikely to be large enough to be measurable in electronic systems using current technology.

Note that the extremely weak reservoir coupling to the bound state will cut-off the bound state-induced correlations on a timescale of order $1/\eta$. In physical systems, this cut-off is the timescale of inelastic scattering. At low temperatures this can be orders of magnitude longer than the decay in the absence of the bound state.

Most results in this work were derived using Heisenberg equations of motion by one of us (E.J.), and then confirmed using the Keldysh approach in Ref.~[\onlinecite{Hasegawa2017Dec}] by M.H. However, the inverse is true for the results in this appendix. Both theoretical approaches can be used for both initial conditions (initial quench and adiabatic preparation), confirming that the discrepancy of Eq.~\eqref{j^(N)_steady} with Meir and Wingreen's result is purely due to different initial conditions. In real experiments, we can expect the initial preparation to be somewhere between an initial quench and adiabatic preparation, so we can expect some oscillations in the dot occupation and currents, but with a smaller magnitude than in Eq.~\eqref{j^(N)_steady}.


\bibliography{ref}

\begin{thebibliography}{60}
\bibitem [{hba()}]{hbar-note}%
  \BibitemOpen
  \href@noop {} {}\bibinfo {note} {We measure energies in units of inverse
  time, so $\hbar=1$.}\BibitemShut {Stop}%
\bibitem [{equ()}]{equilib-note}%
  \BibitemOpen
  \href@noop {} {}\bibinfo {note} {If the two reservoirs have the same $T$ and
  $\mu$, this steady state is the thermal state of the full Hamiltonian in
  Eq.~(\ref{H}).}\BibitemShut {Stop}%
\end{thebibliography}

\end{document}